\documentclass[aps,prc,preprint,amsmath,amssymb,nofootinbib,superscriptaddress,longbibliography,floatfix,raggedbottom]{revtex4-2}

\usepackage{graphicx}
\usepackage{dcolumn}
\newcolumntype{d}[1]{D{.}{.}{#1}}

\graphicspath{{fig/}}
\newcommand{\nubar}{\bar\nu}
\newcommand{\GeV}{\,\mathrm{GeV}}
\newcommand{\MeV}{\,\mathrm{MeV}}
\newcommand{\fm}{\,\mathrm{fm}}
\newcommand{\cm}{\,\mathrm{cm}}
\newcommand{\Lam}{\Lambda}
\newcommand{\Sig}{\Sigma}
\newcommand{\Sst}{\Sigma^{*}(1385)}
\newcommand{\DS}{|\Delta S|}

\providecommand{\sla}[1]{\displaystyle{\not}#1}
\providecommand{\slashed}[1]{{\not}#1}

\begin{document}

\title{Strangeness-changing two-body weak currents\\ from SU(3) chiral dynamics}

\author{J.~A.~Nowak}
\affiliation{School of Physics and Astronomy, Lancaster University, Lancaster LA1 4YB, United Kingdom}

\date{\today}

\begin{abstract}
\begin{description}
\item[Background] Two-particle--two-hole excitations are an established part of neutrino--nucleus scattering, but we know of no microscopic calculation of the semileptonic $\DS=1$ two-body current. Liquid-argon experiments are collecting the antineutrino hyperon samples it feeds.
\item[Purpose] To construct the tree-level strangeness-changing two-body current for $\nubar\,[NN]\to\ell^+ Y N'$ from vertices of the lowest-order SU(3) chiral Lagrangian, test its vector and axial Ward identities, and evaluate its operator families, hyperon composition and charge channels on a correlated pair.
\item[Method] Contact, meson-in-flight, baryon-pole and kaon-exchange families carry symbolically computed flavour-route coefficients, and both line assignments of a proton--neutron pair are summed coherently; direct and crossed decuplet poles carry width-derived couplings, the $N\!\to\!\Delta$ transition form factors and SU(3) route signs. The undressed vector current is the Noether current of the flavour rotation. The axial current $A_{\rm comp}$ carries a vertex-level kaon-pole longitudinal completion fixed by the Goldberger--Treiman relation; the axial contact operator, its kaon-pole partner and the remaining inverse-propagator and mass-splitting terms are omitted. The reference form factors break the vector identity (median residual $0.27$); a Ward-compatible dressing preserves it before current-to-transport matching. A factorised benchmark on a free correlated pair fixes ratios but no absolute scale; four prescriptions separate the current from the cascade.
\item[Results] Before dressing and matching, the SU(3)-limit vector identity closes pointwise (median residual $2\times10^{-10}$), and at physical masses the constructed relation $q\!\cdot\!A_{\rm comp}=(m_s+m_u)P_{su}$ holds chain by chain at machine precision. The flavour routes allow a primary kaon-less $\Sig^+n$ final state from a $pp$ pair, unreachable by any one-body weak vertex; antisymmetrised and weighted with the reference pair census, it is $1.13$--$1.20\%$ of the isolated two-body hyperon rate over all seven charge channels. In the $\Lam$ channel the octet baryon poles carry $77\%$ of the incoherent family sum at $0.6\GeV$ but $43$--$58\%$ at $1$--$5\GeV$ ($30$--$88\%$ at $0.6$--$5\GeV$ across prescriptions). The decuplet carries $26$--$48\%$ at $1$--$5\GeV$ and the meson in flight $0.1$--$0.3\%$; the line assignments interfere destructively. $\sigma_\Lam/\sigma_{\Sig^0}$ of the isolated two-body current is $0.81$--$0.87$ times its quasi-elastic value at $1$--$5\GeV$ (central prescription) and $0.67$--$0.95$ across prescriptions; the axial contact operator and its kaon-pole partner, at the Lagrangian coefficient, lower it by $6\%$ at $1\GeV$ and $2\%$ at $3\GeV$, and the Ward-compatible dressing changes it by about $4\%$. A correlation on every meson changes the composition by under $3\%$; a pion-only Landau--Migdal contact leaves kaon exchange unscreened and roughly triples $\sigma_\Lam/\sigma_{\Sig^0}$.
\item[Conclusions] The $KNY$ couplings carry the axial $N\!\to\!Y$ flavour combinations, so a common scale factor across hyperon channels has no support. Only pion-only screening makes the two-body current more $\Lam$-rich than the quasi-elastic one. Above $1\GeV$ the decuplet transition form factors are a leading input. A short-range treatment must address every exchanged meson, the kaon included.
\end{description}
\end{abstract}

\maketitle

\section{Introduction}
\label{sec:intro}

Two-body currents, the 2p2h or meson-exchange-current contribution, are an
established part of the description of neutrino--nucleus scattering in the few-GeV
region~\cite{nieves2011,gil1997,amaro2017,barbaro2016,martini2009,lovato2020,martinez2025,belocchi2025,barbaro2025}.
The microscopic non-strange operator is pion exchange between two nucleons, with the $W$
attached to the exchanged pion, to the $\pi NN$ vertex or to an intermediate
$\Delta(1232)$, and two nucleons in the final state. We are not aware of a microscopic
calculation of its semileptonic $\DS=1$ counterpart, in which the external weak current
makes a hyperon through the same two-body mechanism; the hadronic $\DS=1$ two-baryon
vertex of hypernuclear decay is discussed in Sec.~\ref{sec:nmwd}.

Antineutrino charged-current hyperon production, $\nubar N\to\ell^+ Y$, is Cabibbo
suppressed but clean: the $\Delta S=\Delta Q$ rule forbids the neutrino-induced channel
and the GIM mechanism the neutral-current one, so a hyperon accompanied by no kaon is an
antineutrino signature~\cite{singh2006,thorpe2021,fatima2019,benitez2024}. Liquid-argon
detectors have reported the first such measurement~\cite{microboone2023}, the DUNE far
detector will collect larger samples~\cite{dune2020}, and hyperon samples probe the
hyperon--nucleus potential and through it the equation of state of dense
matter~\cite{companionPRD,companionPRL}. The $\DS=1$ two-body current feeds the
same samples.

This paper constructs that current at tree level from vertices of the lowest-order
SU(3) chiral Lagrangian: a Kroll--Ruderman contact term, a meson in flight converted
into a kaon by the $K_{\ell3}$ vertex, octet baryon poles in both orderings, a
kaon-exchange family in which the $W$ strikes the partner nucleon, and direct and
crossed decuplet poles with width-derived couplings and the $N\!\to\!\Delta$ transition
form factors, carried to the $\Sst$ by SU(3). Every octet flavour-route coefficient is
computed symbolically, and for a proton--neutron pair the two line assignments are
summed coherently with the relative sign of Fermi statistics. It is a tree-level
chiral model with a restricted operator basis: the axial current carries its
longitudinal kaon-pole completion but omits the axial contact operator and its
kaon-pole partner (Secs.~\ref{sec:constraints} and~\ref{sec:ward}). The
operators are evaluated on a free correlated pair in a factorised approximation, which
fixes ratios between classes and between hyperon channels but no absolute scale; the
matching to the non-strange two-body strength, and the normalisation it gives, are the
subject of Ref.~\cite{Nowak:2026pheno}.

Four results follow. The flavour routes open a primary, kaon-less $\Sig^+$ from a $pp$
pair, which survives antisymmetrisation of the identical protons and is
$1.1$--$1.2\%$ of the two-body hyperon rate (Sec.~\ref{sec:channels}). The flavour
structure does not support a channel-independent rescaling of the quasi-elastic
composition: $\sigma_\Lam/\sigma_{\Sig^0}$ of the two-body current is $0.81$--$0.87$
times its quasi-elastic value at $1$--$5\GeV$ at the central prescription
(Sec.~\ref{sec:composition}). The decuplet carries $26$--$48\%$ of the $\Lam$-channel
family sum at $1$--$5\GeV$, against $6\%$ at $0.6\GeV$, so its transition form factors
are a leading input above $1\GeV$ (Sec.~\ref{sec:hierarchy}). And a pion-only
Landau--Migdal contact~\cite{oset1982,nieves2011} leaves the kaon exchange unscreened
and roughly triples $\sigma_\Lam/\sigma_{\Sig^0}$, whereas a correlation on every
exchanged meson changes the composition by under $3\%$ (Sec.~\ref{sec:screening}).

\section{Kinematics}
\label{sec:xsec}

We consider
\begin{equation}
  \nubar_\ell(k) + A \;\longrightarrow\; \ell^+(k') + X,
  \label{eq:reaction}
\end{equation}
with $q=k-k'$ the four-momentum transfer, $Q^2=-q^2>0$, $\omega=q^0$, and $M_\ell$ the
charged-lepton mass. The strangeness-changing charged-current interaction is governed
at low energies by
\begin{equation}
  \mathcal{H}_W = \frac{G_F}{\sqrt2}\,V_{us}\;
    \ell^\mu\,J^{(s)}_\mu + \mathrm{h.c.},
  \label{eq:hamiltonian}
\end{equation}
with $V_{us}=\sin\theta_C$, $G_F=1.166\times10^{-5}\GeV^{-2}$ (physical-constants table
of Ref.~\cite{pdg2024}), and $\ell^\mu$ the leptonic current. The hadronic current
$J^{(s)}_\mu$ carries $\Delta S=\Delta Q=-1$, which restricts
Eq.~\eqref{eq:reaction} to antineutrinos and forbids the neutral-current analogue by
the GIM mechanism.

The inclusive double-differential cross section is
\begin{equation}
  \frac{d^2\sigma}{d\Omega'\,dE'} =
    \frac{G_F^2\,|V_{us}|^2}{4\pi^2}\,\frac{|\mathbf{k}'|}{|\mathbf{k}|}\;
    L_{\mu\nu}\,W^{\mu\nu},
  \label{eq:xsec}
\end{equation}
where the leptonic tensor for an incoming antineutrino is
\begin{equation}
  L_{\mu\nu} = k_\mu k'_\nu + k'_\mu k_\nu - g_{\mu\nu}(k\!\cdot\!k')
               \;-\; i\,\epsilon_{\mu\nu\alpha\beta}k^\alpha k'^\beta,
  \label{eq:lepton}
\end{equation}
with metric $g=\mathrm{diag}(+,-,-,-)$, $\epsilon^{0123}=-1$ and the antisymmetric term of
the opposite sign for neutrinos; it is $\tfrac18\sum(\ell_\mu)^*\ell_\nu$ for
$\ell^\mu=\bar v_\nu(k)\gamma^\mu(1-\gamma_5)v_\ell(k')$, the factor $8$ of the $V\!-\!A$
trace being in the prefactor of Eq.~\eqref{eq:xsec}. The hadronic tensor is
\begin{equation}
  W^{\mu\nu} = \frac{1}{2M_i}\,\overline{\sum_i}\sum_f (2\pi)^3\,
    \delta^4(P_i+q-P_f)\,
    \langle f|J^{(s)\mu}(0)|i\rangle^{*}\langle f|J^{(s)\nu}(0)|i\rangle ,
  \label{eq:hadronic}
\end{equation}
with $M_i$ the target mass, states normalised covariantly,
$\langle p'|p\rangle=2E\,(2\pi)^3\delta^3(\mathbf p'-\mathbf p)$, the sum over $f$
including the final-state phase-space integrals and the bar denoting the average over
initial spins; this normalisation is that of Ref.~\cite{nieves2011}, and with it
Eq.~\eqref{eq:xsec} is the cross section per target. The order of the indices in
Eq.~\eqref{eq:hadronic} matches that of Eq.~\eqref{eq:lepton}, so that
$L_{\mu\nu}W^{\mu\nu}$ contains $\tfrac18\sum|\ell_\mu\langle f|J^{(s)\mu}|i\rangle|^2$.

The standard decomposition of the inclusive cross section into nuclear response
functions follows Ref.~\cite{amaro2017}; the present work evaluates the current
operators and their matrix elements on a pair, not those responses.

The current decomposes into one- and two-body pieces,
\begin{equation}
  J^{(s)\mu} = \sum_{i} j^{\mu}_{(1)}(i) + \sum_{i<j} j^{\mu}_{(2)}(i,j) + \dots,
  \label{eq:decomposition}
\end{equation}
and the one-body term produces the quasi-elastic hyperon peak; the two are the one- and
two-particle--hole cuts of the same $W$ self-energy (Sec.~\ref{sec:manybody}), and the
separation is exact only in the absence of interference (Sec.~\ref{sec:interference}).

\section{One-body current}
\label{sec:onebody}

For a single nucleon the matrix element of $J^{(s)}_\mu$ between a nucleon of momentum
$p$ and a hyperon of momentum $p'$ takes the standard form
\begin{align}
  \langle Y(p')|J^{(s)\mu}|N(p)\rangle
   = \bar u_Y(p')\Big[& f_1(Q^2)\gamma^\mu
     + \frac{i f_2(Q^2)}{M_N+M_Y}\sigma^{\mu\nu}q_\nu \nonumber\\
    &+ g_1(Q^2)\gamma^\mu\gamma_5
     + \frac{g_3(Q^2)}{M_N+M_Y}q^\mu\gamma_5\Big]u_N(p),
  \label{eq:onebody}
\end{align}
with the second-class axial form factor $g_2$ set to zero. Cabibbo theory fixes the
couplings at $Q^2=0$ in terms of the nucleon electromagnetic form factors and the SU(3)
axial parameters $D$ and $F$~\cite{csw2003}; we use $D=0.804$, $F=0.467$, so that
$D+F=1.271$, within a few parts in a thousand of the measured $g_A$. Table~\ref{tab:su3}
collects them.

\begin{table}[b]
\caption{\label{tab:su3}SU(3) couplings of the one-body $\DS=1$ current at $Q^2=0$,
with $\kappa_p=1.7928$, $\kappa_n=-1.9130$, $D=0.804$, $F=0.467$. The last two columns
give the magnetic combination and the product that fixes the sign of the antineutrino
vector--axial interference.}
\begin{ruledtabular}
\begin{tabular}{lccccc}
transition & $f_1(0)$ & $f_2(0)$ & $g_1(0)$ & $f_1+f_2$ & $g_1(f_1+f_2)$ \\
\hline
$p\to\Lam$    & $\sqrt{3/2}$   & $\sqrt{3/2}\,\kappa_p$      & $-(D+3F)/\sqrt6$ & $3.420$ & $-3.079$ \\
$p\to\Sig^0$  & $1/\sqrt2$     & $(\kappa_p+2\kappa_n)/\sqrt2$ & $(D-F)/\sqrt2$ & $-0.731$ & $-0.174$ \\
$n\to\Sig^-$  & $1$            & $\kappa_p+2\kappa_n$      & $D-F$ & $-1.033$ & $-0.348$ \\
\end{tabular}
\end{ruledtabular}
\end{table}

The $Q^2$ dependence of the form factors is dipole, with $M_V=0.84\GeV$ and $M_A\simeq1.0\GeV$~\cite{bernard2002}.

The induced pseudoscalar $g_3$ follows from the axial Ward identity. Contracting the
current of Eq.~\eqref{eq:onebody} with $q_\mu$ and using
$\bar u_Y\slashed q\gamma_5u_N=(M_N+M_Y)\bar u_Y\gamma_5u_N$ gives
\begin{equation}
  q_\mu A^\mu_{NY}
  = \left[(M_N+M_Y)g_1(q^2)+\frac{q^2}{M_N+M_Y}g_3(q^2)\right]\bar u_Y\gamma_5 u_N .
  \label{eq:ward1b}
\end{equation}
With the kaon-pole form of the induced pseudoscalar demanded by PCAC,
\begin{equation}
  g_3(q^2) = -\frac{(M_N+M_Y)^2}{q^2-m_K^2}\,g_1(q^2)
           = \frac{(M_N+M_Y)^2}{m_K^2+Q^2}\,g_1(q^2),
  \label{eq:g3pcac}
\end{equation}
the bracket collapses to $(M_N+M_Y)g_1\,m_K^2/(m_K^2-q^2)$.

\section{Two-body current}
\label{sec:twobody}

\subsection{Vertices}
\label{sec:lagrangian}
\label{sec:su3}

We work with the lowest-order chiral Lagrangian~\cite{gss1988,jenkins1991,bkm1995} for
the ground-state baryon octet $B$ and the pseudoscalar octet $\Phi$,
\begin{equation}
  \mathcal{L} = \langle \bar B\,(i\sla D - M_B)B\rangle
    + \frac{D}{2}\langle\bar B\gamma^\mu\gamma_5\{u_\mu,B\}\rangle
    + \frac{F}{2}\langle\bar B\gamma^\mu\gamma_5[u_\mu,B]\rangle,
  \label{eq:lagrangian}
\end{equation}
with $u_\mu = i\{\xi^\dagger\partial_\mu\xi - \xi\partial_\mu\xi^\dagger\}$,
$\xi^2=U=\exp(i\sqrt2\Phi/f)$, and the external charged weak current introduced
through the covariant derivative $D_\mu$ and through $u_\mu$. The strong meson--baryon
couplings are
\begin{align}
  g_{\pi NN} &= \frac{D+F}{2f_\pi} = \frac{g_A}{2f_\pi}, \label{eq:gpinn}\\
  g_{K N\Lam} &= -\frac{D+3F}{2\sqrt3\,f_\pi}, \label{eq:gknl}\\
  g_{K N\Sig} &= \frac{D-F}{2f_\pi}, \label{eq:gkns}
\end{align}
with $f_\pi = 92.4\MeV$. These are pseudovector couplings, of dimension inverse mass;
$2M_N\,g_{\pi NN}=M_Ng_A/f_\pi=12.9$ is the conventional dimensionless $\pi NN$
coupling. We work in the SU(3)-symmetric limit $f_K=f_\pi$, in which the
Goldberger--Treiman~\cite{goldberger1958} reduction of Sec.~\ref{sec:constraints} holds;
the physical $f_K\simeq1.19f_\pi$ is a next-order effect not included.

The weak vertices we need are the $K$ and $\pi$ decay constants, in the normalisation
$f_\pi=92.4\MeV$,
\begin{equation}
  \langle 0|J^{(s)}_\mu|K(p)\rangle = i\sqrt2\,f_K p_\mu,
  \label{eq:fk}
\end{equation}
and the strangeness-changing meson transition current
\begin{equation}
  \langle K(p_K)|J^{(s)}_\mu|\pi(p_\pi)\rangle
    = f_+(t)\,(p_K+p_\pi)_\mu + f_-(t)\,(p_K-p_\pi)_\mu,
  \label{eq:fplus}
\end{equation}
with $t=(p_K-p_\pi)^2$ and the form factors of $K_{\ell3}$ decay, $f_+(0)=0.970$ and
the measured linear slope $\lambda_+=0.0297$~\cite{pdg2024} (Table~\ref{tab:inputs}),
kept as a convention and not a controlled spacelike extrapolation; the class carries
$0.1$--$0.3\%$ of the incoherent sum (Sec.~\ref{sec:hierarchy}), which bounds the
effect of the form-factor shape.
Equation~\eqref{eq:fplus} is the strange analogue of the vertex that generates the
pion-in-flight current in the non-strange case: the weak current changes the
strangeness of the exchanged meson without touching either baryon line.

\subsection{Operator classes}
\label{sec:classes}

\begin{figure*}[t]
\centering
\includegraphics[width=\textwidth]{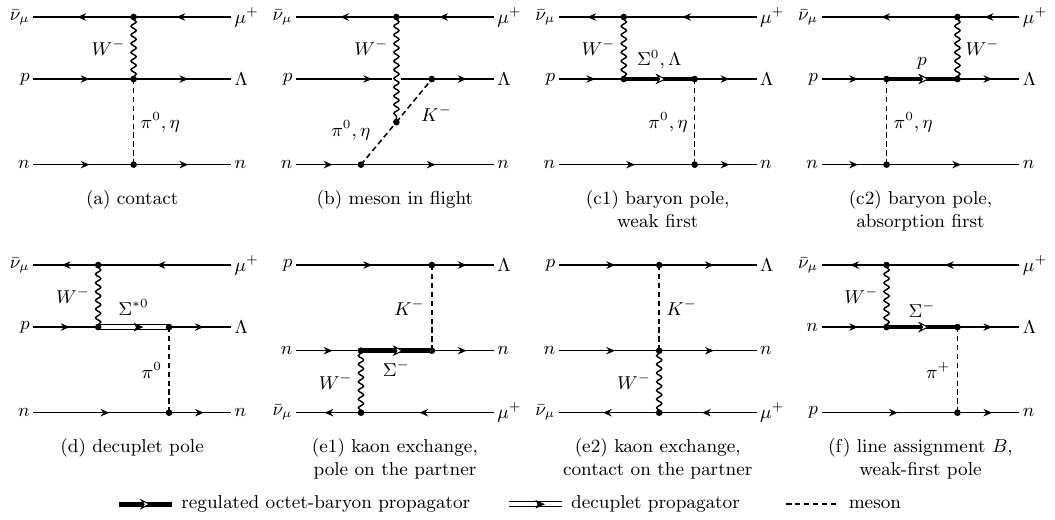}
\caption{\label{fig:topologies}Topologies of the $\DS=1$ two-body current, drawn for
$\nubar_\mu\,(pn)\to\mu^+\Lam n$. Panels (a)--(e) are the classes of
Sec.~\ref{sec:classes} in line assignment $A$ (proton on line 1), each on its
canonical route: (a) contact term, Eq.~\eqref{eq:contact}; (b) meson-in-flight term,
Eq.~\eqref{eq:inflight}; (c1) weak-first and (c2) absorption-first baryon poles,
Eqs.~\eqref{eq:hypole} and~\eqref{eq:nucpole}; (d) direct decuplet pole,
Eq.~\eqref{eq:decuplet}, the crossed and line-2 decuplet terms not drawn; (e1) and
(e2) the kaon-exchange family, with the $W$ on the partner nucleon, as a hyperon pole
and as a contact term, Eq.~\eqref{eq:kexch}. Panel (f) is the weak-first pole of line
assignment $B$ (neutron on line 1) in Eq.~\eqref{eq:fullcurrent}. Thick segments with an
open arrowhead are octet-baryon propagators regulated by the matching prescription; the
double line is the decuplet propagator.
Each meson line is labelled with the meson exchanged.}
\end{figure*}

The two nucleons exchange a meson, and the external $W$ attaches to the meson line, to
a meson--baryon vertex, to a baryon line, or to an intermediate resonance. Strangeness
must be created somewhere, which fixes the identity of the exchanged meson in each
class. We label the two nucleons $1$ and $2$ and take nucleon $1$ to become the hyperon,
$N_1(p_1)\,N_2(p_2)\to Y(p_1')\,N'(p_2')$. Every exchanged meson flows from line 2 to
line 1, a pion or $\eta$ with momentum $k_\pi=p_2-p_2'$ and a kaon with
$k_K=k_\pi+q=p_1'-p_1$. A common factor of $i$ is omitted from every amplitude below;
the relative signs of the octet terms are those of the insertion sum of
Eq.~\eqref{eq:noether3}. Figure~\ref{fig:topologies}
draws each class for $\nubar_\mu\,(pn)\to\mu^+\Lam n$.

Every class shares the partner-line factor in which nucleon 2 emits the exchanged meson
$\phi$ through the pseudovector strong vertex and recoils,
\begin{equation}
  L^{\phi}_{2} \;=\; \frac{\sqrt2\,c^{\phi}_{2}}{f}\;
    \widetilde D_\phi(k_\pi)\;
    \bar u_{N'}\sla k_\pi\gamma_5\,u_{N_2},
  \qquad
  \widetilde D_\phi(k) = \frac{F_\phi(k^2)}{k^2-m_\phi^2},
  \label{eq:line2}
\end{equation}
with $c^{\phi}_{2}$ the flavour trace of Sec.~\ref{sec:isospin} for that emission and
$F_\phi$ the monopole form factor of Table~\ref{tab:inputs}, one per exchanged-meson
propagator. Each class below is a line-1 structure multiplying $L^\phi_2$, summed
coherently over $\phi$ and over the charge routes; the coefficients $c^\phi_i$ are
flavour traces throughout, and with the factor $\sqrt2/f$ per vertex they reproduce the
physical couplings of Eqs.~\eqref{eq:gpinn}--\eqref{eq:gkns}
(Sec.~\ref{sec:isospin}).

\paragraph*{(a) Contact term.} Expanding $u_\mu$ in Eq.~\eqref{eq:lagrangian} to one
meson and one external current, $u_\mu\supset-(i\sqrt2/f)[\Phi,v_\mu]$, generates a
Kroll--Ruderman seagull in which the $W$, the exchanged meson and both baryons meet
at a single vertex on the hyperon line:
\begin{equation}
  j^\mu_{(a)} = \frac{\sqrt2\,c^{\phi}_{\rm sg}}{f}\;
    \bar u_Y\gamma^\mu\gamma_5\,u_{N_1}\;\times\;L^{\phi}_{2},
  \label{eq:contact}
\end{equation}
where $c^\phi_{\rm sg}$ is the seagull trace of Eq.~\eqref{eq:traces}, on the $\Lam$ channel $-(D{+}3F)/(4\sqrt3)$ for
$\pi^0$ exchange and $-(D{+}3F)/4$ for $\eta$. The exchanged meson is a pion (or
$\eta$), the strangeness change being carried by the contact vertex, so this class
retains the long range $1/m_\pi$ of the non-strange seagull. In the reference
evaluation the seagull carries the monopole of $L^\phi_2$ and no weak form factor.

\paragraph*{(b) Meson-in-flight.} Nucleon 2 emits the meson through the ordinary
strong vertex; the weak current converts it into a kaon through
Eq.~\eqref{eq:fplus}; the kaon is absorbed on nucleon 1, making the hyperon:
\begin{equation}
  j^\mu_{(b)} = \frac{\sqrt2\,c^{\phi}_{\rm fl}}{f}\,
    f_+(t)\,\frac{F_K(k_K^2)}{k_K^2-m_K^2}\,(k_K+k_\pi)^\mu\;
    \bar u_Y\sla k_K\gamma_5\,u_{N_1}\;\times\;L^{\phi}_{2},
  \label{eq:inflight}
\end{equation}
where the $K_{\ell3}$ vertex carries $f_+(t)$ and no further form factor. The
$f_-$ term of Eq.~\eqref{eq:fplus}, which multiplies $q^\mu$ and changes the rate of
every channel by at most $0.03\%$, is omitted.
This class carries two meson propagators, one of them a kaon, so its range is set by
$1/m_K\simeq0.40\fm$.

\paragraph*{(c) Baryon-pole terms.} The weak current acts on one baryon line through
the one-body vertex of Eq.~\eqref{eq:onebody}, and the resulting baryon exchanges a
meson with the partner. Two orderings contribute. The weak transition first gives an
intermediate hyperon $Y_v$, summed over the charge routes with its own mass and weak
vertex:
\begin{equation}
  j^\mu_{(c1)} = \frac{\sqrt2\,c^{\phi}_{\rm yr}}{f}\;
    \frac{\bar u_Y \sla k_\pi\gamma_5\,
      \big(\sla p_1+\sla q + M_{Y_v}\big)\,\Gamma^\mu_{N\to Y_v}\,u_{N_1}}
      {(p_1+q)^2-M_{Y_v}^2}\;\times\;L^{\phi}_{2},
  \label{eq:hypole}
\end{equation}
and the strong exchange first gives an intermediate nucleon,
\begin{equation}
  j^\mu_{(c2)} = \frac{\sqrt2\,c^{\phi}_{\rm np}}{f}\;
    \frac{\bar u_Y \Gamma^\mu_{N\to Y}\,
      \big(\sla p_1+\sla k_\pi + M_N\big)\,\sla k_\pi\gamma_5\,u_{N_1}}
      {(p_1+k_\pi)^2-M_N^2}\;\times\;L^{\phi}_{2},
  \label{eq:nucpole}
\end{equation}
with $\Gamma^\mu_{NY}$ the bracket of Eq.~\eqref{eq:onebody} carrying the propagating
pseudoscalar of Eq.~\eqref{eq:properpole}. In both, $1/\mathrm{den}$ stands for the
regulated propagator of Sec.~\ref{sec:boundary}: zero for $|\mathrm{den}|\le\Delta$ in
the formation-length cut, $\mathrm{den}/(\mathrm{den}^2+\epsilon^2)$ in the
regulated principal-value model, the numerators being unchanged. These are the terms called
``correlation'' diagrams in the non-strange language, and the ones most at risk of
double counting against a final-state-interaction treatment of the one-body process
(Sec.~\ref{sec:interference}).

\paragraph*{(d) Decuplet-pole terms.} The intermediate baryon may belong to the decuplet.
In the direct term the weak current excites a $\Sst$ on line 1, which absorbs the
exchanged meson and leaves as the final hyperon; with $\mathcal{P}^{\alpha\beta}$ the
Rarita--Schwinger propagator of Eq.~\eqref{eq:rs} and $\Gamma^{\beta\mu}_{N\to\Sst}$ the
weak transition vertex,
\begin{equation}
  j^\mu_{(d)} = \frac{g_{\Sst Y\pi}}{m_\pi}\,
    \bar u_Y\, k_\pi^\alpha\,
    \mathcal{P}_{\alpha\beta}(p_1+q)\,
    \Gamma^{\beta\mu}_{N\to\Sst}\,u_{N_1}\;\times\;L^{\phi}_{2}.
  \label{eq:decuplet}
\end{equation}
In the crossed term line 1 first absorbs the pion into a $\Delta(1232)$, and the weak
current converts the $\Delta$, of momentum $p_1+k_\pi=p_1'-q$, into the final hyperon
through the conjugate vertex $\gamma^0[\Gamma^{\beta\mu}(p_1',-q)]^\dagger\gamma^0$. In
the topology of class~(e) the $W$ excites a $\Sst$ on line 2, which emits the kaon
absorbed on line 1. The class is not generated by the octet
Lagrangian of Eq.~\eqref{eq:lagrangian}: its strong couplings are fixed from measured
widths, its weak vertex from the $N\!\to\!\Delta$ transition, and its route structure
from SU(3). The Rarita--Schwinger propagator is
\begin{equation}
  \mathcal{P}^{\alpha\beta}(p) = \frac{\slashed p + M_R}{p^2-M_R^2+iM_R\Gamma_R}\,
    P^{\alpha\beta}(p),
  \label{eq:rs}
\end{equation}
\begin{align}
  P^{\alpha\beta} = -g^{\alpha\beta}
    &+ \tfrac13\gamma^\alpha\gamma^\beta
     + \tfrac{1}{3M_R}\!\left(\gamma^\alpha p^\beta - \gamma^\beta p^\alpha\right)
     \nonumber\\
    &+ \tfrac{2}{3M_R^2}p^\alpha p^\beta .
  \label{eq:rsproj}
\end{align}
The weak vertex is that of Ref.~\cite{hnv2007},
\begin{align}
  \Gamma^{\beta\mu} = \Big[&\frac{C_3^V}{M}\big(g^{\beta\mu}\slashed q-q^\beta\gamma^\mu\big)
    +\frac{C_4^V}{M^2}\big(g^{\beta\mu}q\!\cdot\!p_R-q^\beta p_R^\mu\big) \nonumber\\
   &+\frac{C_5^V}{M^2}\big(g^{\beta\mu}q\!\cdot\!p-q^\beta p^\mu\big)\Big]\gamma_5
    + C_5^A\,g^{\beta\mu},
  \label{eq:decvertex}
\end{align}
with $M$ the nucleon mass and $p$ and $p_R=p+q$ the nucleon and resonance momenta. The
vector part is transverse term by term; the axial part carries no $\gamma_5$ and is
completed by its constant-free $C_6$ partner (Sec.~\ref{sec:constraints}); $C_4^A$ is
omitted. The $\Sst$ vertices take the same form factors times the SU(3) transition
strengths relative to $p\!\to\!\Delta^{++}$~\cite{alam2012}. The strong $R\!\to\!B\pi$
vertex is $(g/m_\pi)\bar u_B k^\alpha u_{R,\alpha}$ with $g$ fixed from the partial width
summed over charge states,
\begin{equation}
  \Gamma_{R\to B\pi} = \frac{1}{6\pi}\left(\frac{g}{m_\pi}\right)^2
     \frac{M_B}{M_R}\,|\mathbf{k}|^3 ,
  \label{eq:width}
\end{equation}
the static approximation of the Rarita--Schwinger width. The couplings, widths and form
factors are listed in Table~\ref{tab:inputs}, the SU(3) transition strengths and isospin
factors of the routes follow Refs.~\cite{hnv2007,alam2012,pdg2024}, and the phase of
the class relative to the octet families is fixed by the non-strange
Goldberger--Treiman relations for $g_A$ and $C_5^A$~\cite{hnv2007}.

\paragraph*{(e) Kaon-exchange family.} One family does not factor through
$L^\phi_2$. The $W$ strikes the partner, whose intermediate hyperon emits a $\bar K$
that crosses and converts line 1. The common factor is a kaon absorption on line 1,
\begin{equation}
  L_1 \;=\; \frac{\sqrt2\,a_1}{f}\,\widetilde D_K(k_K)\;
    \bar u_Y\sla k_K\gamma_5\,u_{N_1},
  \label{eq:line1}
\end{equation}
and the weak chain runs on line 2 with intermediate momentum $p_2+q$:
\begin{equation}
  j^\mu_{(e)} = \Big[\frac{\sqrt2\,c_{\rm r}}{f}\,
    \frac{\bar u_{N'}\sla k_K\gamma_5\big(\sla p_2+\sla q+M_{Y_v}\big)
      \Gamma^\mu_{N\to Y_v}u_{N_2}}{(p_2+q)^2-M_{Y_v}^2}
    \;-\;\frac{\sqrt2\,c_{\rm sg2}}{f}\,\bar u_{N'}\gamma^\mu\gamma_5 u_{N_2}\Big]\,L_1 ,
  \label{eq:kexch}
\end{equation}
the relative minus on the line-2 seagull being fixed by the insertion sum of
Eq.~\eqref{eq:noether3}. The
hyperon pole carries the regulated propagator of Eqs.~\eqref{eq:hypole}
and~\eqref{eq:nucpole}. Only the weak-first ordering occurs, since no $\DS=0$ weak
vertex can close the absorption-first one. Its exchanged meson is a kaon, so a
pion-only screening such as Eq.~\eqref{eq:vl} leaves it untouched, while the
correlation function of Eq.~\eqref{eq:src} acts on it like any other exchange
(Sec.~\ref{sec:screening}).

\paragraph*{The completed current.} Equations~\eqref{eq:contact}--\eqref{eq:decuplet}
and~\eqref{eq:kexch} display each class on its canonical route. The evaluated current is the coherent sum
over the full route structure of Sec.~\ref{sec:isospin},
\begin{align}
  j^\mu \;=\; \sum_{a\,=\,A,B}\eta_a\bigg[\;&\sum_{\phi\,=\,\pi,\eta}\;\sum_{\text{routes}\,\in\,a}
  \Big( j^\mu_{\rm sg} + j^\mu_{\rm IF} + j^\mu_{{\rm BP},1} + j^\mu_{{\rm BP},2}
        + j^\mu_{\rm dec} \Big)_{\phi,\,\text{route}} \nonumber\\
  &+\; \sum_{\text{routes}\,\in\,a} j^\mu_{\rm KX}\bigg],
  \label{eq:fullcurrent}
\end{align}
where $a$ labels the two line assignments of a $pn$ pair, $\eta_A=1$, $\eta_B=-1$,
$j^\mu_{\rm dec}$ contains the direct and crossed decuplet terms, and $j^\mu_{\rm KX}$
contains the line-2 $\Sst$, which the family decompositions of Sec.~\ref{sec:numresults}
count with the decuplet. In $A$ the weak current acts on the nucleon that the naive
charge rule turns into the final hyperon, the proton for $\Lam$ and $\Sig^0$ and the
neutron for $\Sig^-$; in $B$ it acts on the other nucleon, and the final state is reached
through charged-pion routes (for $pn\to\Lam n$, $n\to\Sig^-$ at the weak vertex followed
by $\Sig^-\pi^+\to\Lam$ with $p\to n\pi^+$ on the partner). Assignment $B$ is evaluated
with the two initial nucleons' momenta and spins exchanged and enters with the relative
minus sign of the Wick contraction, whether or not the nucleons are identical;
distinguishability removes only the symmetry factor of Eq.~\eqref{eq:2p2h}.

\subsection{Flavour structure}
\label{sec:isospin}

A struck proton yields $\Lam$ or $\Sig^0$ at the weak vertex and a struck neutron
$\Sig^-$, but the final hyperon need not be the one the weak vertex made: a $\Sig^-$
created at the weak vertex can absorb a $\pi^+$ from the partner and leave as a $\Lam$.
Such charge-exchange routes reach the same final state from the same pair and add
coherently; strangeness can also flow through the exchange itself (the kaon-exchange
family), and $\eta$ exchange accompanies $\pi^0$ exchange wherever the flavour traces
allow it. For a $pn$ pair every final state is reached through both line assignments,
the charged-pion routes being those in which the weak current acts on the nucleon
without the naive label: $\Lam$ and $\Sig^0$ from a struck neutron through
$\Sig^-\pi^+$, $\Sig^-$ from a struck proton through $\Lam\pi^-$ and $\Sig^0\pi^-$.

Charge, strangeness and $\Delta S=\Delta Q$ at the weak vertex permit the channels of
Table~\ref{tab:channels} (the overall $pp\to\Sig^+n$ transition also has $\Delta
Q=\Delta S=-1$). The pion is absorbed on line 1 after the weak vertex (weak-first
ordering) or before it (absorption-first, e.g.\ $n\pi^+\!\to\!p$), so a charged pion
moves one unit of charge between the lines. $pp\to\Sig^+n$ has the weak-first ordering
only: $p\pi^+$ has charge $2$, which no octet baryon carries.

\begin{table}[b]
\caption{\label{tab:channels}Channels allowed by charge, strangeness and $\Delta S=\Delta
Q$ at the weak vertex, labelled by the nucleon $N_1$ that becomes the hyperon, the
charge of the exchanged pion and the partner's final state $N'$.}
\begin{ruledtabular}
\begin{tabular}{lllll}
pair & $N_1$ & $\pi$ & $N'$ & final \\
\hline
$pn$ & $p$ & $0$ & $n$ & $\Lam n$, $\Sig^0 n$ \\
$pn$ & $n$ & $+$ & $n$ & $\Lam n$, $\Sig^0 n$ \\
$pn$ & $n$ & $0$ & $p$ & $\Sig^- p$ \\
$pn$ & $p$ & $-$ & $p$ & $\Sig^- p$ \\
$pp$ & $p$ & $0$ & $p$ & $\Lam p$, $\Sig^0 p$ \\
$pp$ & $p$ & $+$ & $n$ & $\mathbf{\Sig^+ n}$ \\
$nn$ & $n$ & $0$ & $n$ & $\Sig^- n$ \\
\end{tabular}
\end{ruledtabular}
\end{table}

Every flavour coefficient is computed from the traces of Sec.~\ref{sec:lagrangian},
with the $\DS=1$ source $h$ the matrix taking a $u$ quark to an $s$:
\begin{align}
  \text{one body}  &: \tfrac{D}{2}\langle\bar B\{h,B\}\rangle
                    + \tfrac{F}{2}\langle\bar B[h,B]\rangle, \nonumber\\
  \text{strong}    &: \tfrac{D}{2}\langle\bar B\{\phi,B\}\rangle
                    + \tfrac{F}{2}\langle\bar B[\phi,B]\rangle, \label{eq:traces}\\
  \text{seagull}   &: \tfrac{D}{2}\langle\bar B\{[h,\phi],B\}\rangle
                    + \tfrac{F}{2}\langle\bar B[[h,\phi],B]\rangle, \nonumber
\end{align}
the seagull structure being the commutator the source expansion generates
($u_\mu\supset-\tfrac{i\sqrt2}{f}[\Phi,v]$); the pole classes are products of the first
two with a propagator between. With $U=\exp(i\sqrt2\Phi/f)$ every octet vertex and
seagull of Eq.~\eqref{eq:lagrangian} is $\sqrt2$ times its trace over $f$, a factor
that cancels in the ratios between octet routes but not against the decuplet, whose
couplings come from the widths and the $N\!\to\!\Delta$ transition.

The exchanged meson is charged in the dominant routes, so the current connects $pn$
pairs preferentially, as in the $\DS=0$ case and in line with the tensor-force dominance
of $pn$ short-range-correlated pairs~\cite{hen2017}; the $pp$ channel opens a direct
$\Sig^+$ (Sec.~\ref{sec:channels}). The symmetry factors of the response are channel
dependent (Eq.~\eqref{eq:2p2h}); the family and composition results of
Sec.~\ref{sec:numresults} are for the $pn$ channels, and the like-pair channels enter in
Sec.~\ref{sec:channels}. A recent microscopic treatment finds that proton--neutron
asymmetry matters at the ten-percent level in $^{40}$Ar relative to
$^{40}$Ca~\cite{martinez2025}.

\subsection{Ward identities}
\label{sec:constraints}

The vector and axial Ward identities are separate statements,
\begin{equation}
  q_\mu V^\mu = 0 \ \ \text{(SU(3) limit)}, \qquad
  q_\mu A^\mu = 0 \ \ \text{(chiral limit)},
  \label{eq:twoids}
\end{equation}
with distinct right-hand sides away from those limits, so the two sectors are built and
contracted independently throughout.

\paragraph{Vector current.} In the SU(3) limit the $\DS=1$ vector current is the
Noether current of the flavour rotation $\delta X=i[h,X]$: the baryon term
$-\langle\bar B\gamma^\mu[h,B]\rangle$, whose insertions on the external legs are the
pole classes on both lines (the line-2 insertions being the kaon-exchange family); the
meson kinetic term $i\langle\partial^\mu\Phi[h,\Phi]\rangle$, the in-flight
class; and the Kroll--Ruderman seagull
$-\tfrac{i\sqrt2}{f}\big[\tfrac{D}{2}\langle\bar B\gamma^\mu\gamma_5\{[h,\Phi],B\}\rangle
+\tfrac{F}{2}\langle\cdots\rangle\big]$ at both strong vertices. Every relative
coefficient is fixed by the one construction, and $q_\mu V^\mu=0$ is an identity.

\paragraph{Scope of the evaluated axial current.} The full off-shell axial
Ward--Takahashi identity for a baryon transition has the schematic form
\begin{equation}
  q_\mu\Gamma_A^\mu
  = S_Y^{-1}(p')\,\gamma_5 T + \gamma_5 T\,S_N^{-1}(p) + (m_s+m_u)\,\Gamma_5,
  \label{eq:wtoffshell}
\end{equation}
with $S^{-1}(p)=\slashed p-M$, $T$ the axial coupling of the transition and $\Gamma_5$ the
vertex of $P_{su}$, defined in the PCAC paragraph below. Its inverse-propagator terms
vanish between on-shell spinors but, inside a two-body chain with one leg off shell, tie
pole and contact contributions together. The axial current of
Eq.~\eqref{eq:lagrangian}, with the next-order operators that split the octet masses, is
\begin{equation*}
  A^\mu = A^\mu_{\rm comp}+A^\mu_{\rm ga}+A^\mu_{{\rm ga},K}+A^\mu_{\rm ip}+A^\mu_{\rm sb},
\end{equation*}
with $A^\mu_{\rm comp}$ the evaluated current, which implements only the longitudinal
kaon-pole part of the identity, vertex by vertex, so that its divergence has the PCAC
explicit-breaking form $(m_s+m_u)P_{su}$; $A^\mu_{\rm ga}$ the axial contact operator
and $A^\mu_{{\rm ga},K}$ its kaon-pole partner, given below; $A^\mu_{\rm ip}$ the
remaining inverse-propagator and contact contributions to Eq.~\eqref{eq:wtoffshell};
and $A^\mu_{\rm sb}$ the terms of the mass-splitting operators. Only $A^\mu_{\rm comp}$
is evaluated; $A^\mu_{\rm ga}$ and $A^\mu_{{\rm ga},K}$ must be restored together, and
Sec.~\ref{sec:composition} restores these two alone. The transverse part,
$g_1\gamma^\mu\gamma_5$ with its dipole, remains phenomenological input.

\paragraph{Longitudinal completion.} The induced pseudoscalar of Eq.~\eqref{eq:g3pcac}
is the on-shell reduction, via $\bar u\slashed q\gamma_5u=(M+M')\bar u\gamma_5u$, of the
propagating structure the source derivation gives,
\begin{equation}
  A^\mu \;\supset\; -\,g_1\,\frac{q^\mu\,\slashed q\gamma_5}{q^2-m_K^2},
  \label{eq:properpole}
\end{equation}
with no free constant: the matching is the Goldberger--Treiman relation. Inside a
two-body chain with one leg off shell the two forms differ, and the propagating form is
used. The axial part of every vertex then satisfies
\begin{equation}
  q_\mu A^\mu_{{\rm comp},\,{\rm vertex}}
  = g_1\,\slashed q\gamma_5\Big(1-\frac{q^2}{q^2-m_K^2}\Big)
  = -\,g_1\,\frac{m_K^2\,\slashed q\gamma_5}{q^2-m_K^2},
\end{equation}
an operator identity: the divergence of each pole diagram is $m_K^2$-proportional at
every off-shell point, without cancellation between diagrams, so
Eq.~\eqref{eq:properpole} replaces the inverse-propagator terms of
Eq.~\eqref{eq:wtoffshell} rather than implementing them. The axial decuplet vertex
receives the same construction, $T^\mu\to T^\mu-q^\mu(q\!\cdot\!T)/(q^2-m_K^2)$, the
Rarita--Schwinger $C_6$ partner of $C_5^A$, with the pion pole for the $\Delta$ of the
non-strange mirror; the vector part of Eq.~\eqref{eq:decvertex} is transverse and needs
none.

\paragraph{Axial contact operator and its kaon-pole partner.} The chiral connection of
the baryon kinetic term generates the one-meson axial contact operator $A^\mu_{\rm ga}$,
of Dirac structure $\gamma^\mu$, at the same chiral order as the vector seagull of
Eq.~\eqref{eq:contact}, and the Weinberg--Tomozawa vertex of the same term its
kaon-pole partner $A^\mu_{{\rm ga},K}$, with $\gamma^\mu$ replaced by
$-q^\mu(\slashed q-\slashed k_x)/[2(q^2-m_K^2)]$, $k_x$ the momentum of the exchanged
meson entering the vertex; both carry the same flavour coefficient and a common
strength $g$, listed in the Supplemental Material~\cite{P10supp}, the Lagrangian of
Eq.~\eqref{eq:lagrangian} having $g=-1$. The contact alone breaks chiral-limit
conservation. Between the external spinors of the contact diagram
$\bar u_Y(\slashed q+\slashed k_x)u_N=(M_Y-M_N)\,\bar u_Yu_N$, so contact and partner
together form the completion $T^\mu\to T^\mu-q^\mu(q\!\cdot\!T)/(q^2-m_K^2)$ of the
contact and restore conservation (Sec.~\ref{sec:ward}), up to an SU(3)-breaking
remnant proportional to $M_Y-M_N$ that the mass-splitting operators of the full
Lagrangian cancel. Both are omitted from $A^\mu_{\rm comp}$; their effect at $g=-1$ is
studied in Sec.~\ref{sec:composition}.

\paragraph{PCAC relation.} At physical masses PCAC predicts
$q_\mu A^\mu_{su}=(m_s+m_u)P_{su}$, with the pseudoscalar density
$P^a=\sqrt2\,F_0B_0\langle T^a\Phi\rangle$ defined from the scalar--pseudoscalar source
$\chi=2B_0(s+ip)$ of the mesonic Lagrangian and normalised by the
Gell-Mann--Oakes--Renner relation, $(m_s+m_u)\langle 0|\hat P_{su}|K^+\rangle=\sqrt2
f_Km_K^2$ with $f_K=F_0$. Between baryon states,
$P_{su}\equiv-i\langle f|\hat P_{su}(0)|i\rangle$ proceeds through the kaon pole and its
strong absorption, so that with $f_K=f$ the $p\to\Lam$ element is
$(m_s+m_u)P_{su}=-2T_s^{K}(\Lam,p)\,m_K^2/(q^2-m_K^2)\;\bar u_\Lam\slashed q\gamma_5u_p$,
with $T_s^{K}(Y,N)$ the flavour trace of the strong $K$-absorption vertex, while the
completed vertex has $q_\mu\Gamma^\mu_A=-g_1\,m_K^2/(q^2-m_K^2)\,\slashed q\gamma_5$. The
two agree if
\begin{equation}
  g_1(N\!\to\!Y) \;=\; 2\,T_s^{K}(Y,N),
  \label{eq:gtreduce}
\end{equation}
the Goldberger--Treiman relation for the $KNY$ couplings, which the symbolic trace
generator verifies with zero remainder:
\begin{equation}
  \begin{aligned}
  p\to\Lambda:&\quad g_1=-\tfrac{D+3F}{\sqrt6}=-0.9002=2T_s^K,\\
  p\to\Sigma^0:&\quad g_1=+\tfrac{D-F}{\sqrt2}=+0.2383=2T_s^K,\\
  n\to\Sigma^-:&\quad g_1=D-F=+0.3370=2T_s^K.
  \end{aligned}
  \label{eq:gtvalues}
\end{equation}
The relation then holds chain by chain in the assembled amplitude; it follows from the
construction of $A^\mu_{\rm comp}$ and is not a test of Eq.~\eqref{eq:wtoffshell}, and
its pointwise evaluation (Sec.~\ref{sec:ward}) checks that both sides carry the same
routing, coefficients, propagators and regulators.

\subsection{Hypernuclear connection}
\label{sec:nmwd}

The hadronic $\DS=1$ two-baryon vertex of hypernuclear decay, $\Lam N\to NN$, is
constrained by one-boson-exchange analyses of the measured decay
rates~\cite{parreno1997,alberico2002,itonaga2018}, of the crossed reaction $pn\to
p\Lam$~\cite{parreno1999} and of the two-nucleon-induced decay~\cite{itonaga2023}. Its
weak couplings do not transfer to the semileptonic current, in which the $W$ carries
four-momentum $q$ into the hadronic system; its strong sector, the $KNY$ and $K^\ast NY$
couplings~\cite{petschauer2020}, checks the input of Sec.~\ref{sec:su3}.

\subsection{Implications}
\label{sec:apriori}

\paragraph{Direct $\Sigma^+$ channel.} The $\Delta S=\Delta Q$ rule forbids a $\Sig^+$
at the weak vertex, but a $\pi^+$ absorbed at the strong vertex of the internal line
converts the vertex hyperon, $\Lam\pi^+\!\to\!\Sig^+$ and $\Sig^0\pi^+\!\to\!\Sig^+$, the
partner proton becoming a neutron, so a $pp$ pair makes a $\Sig^+$ with no kaon in the
primary hadronic state. Its existence follows from the flavour routes alone; the
fraction of the two-body rate it carries depends on the matching prescription, the
axial truncation, the decuplet model, the pair census and the antisymmetrisation of the
two identical protons (Sec.~\ref{sec:channels}).

\paragraph{Range of the meson-in-flight term.} The in-flight class~(b) carries a kaon on
the strange leg, $1/m_K\simeq0.40\fm$ against the $1/m_\pi\simeq1.4\fm$ of the
pion-in-flight term that dominates the non-strange current, and at the exchanged
momenta a correlated pair supplies, $|\mathbf k|\simeq0.4$--$0.8\GeV$, the kaon
propagator is $0.44$--$0.75$ of a pion one; the contact class~(a) keeps the long range
because its strangeness change sits at the vertex. The argument does not order the
contact against kaon exchange, which also carries a kaon propagator, and it presumes a
long-ranged pion, which a pion-only screening removes (Sec.~\ref{sec:screening}).

\section{Nuclear response}
\label{sec:manybody}

\subsection{Hadronic tensor}
\label{sec:selfenergy}

Following Refs.~\cite{gil1997,nieves2011}, the inclusive hadronic tensor of a nucleus
is obtained from the imaginary part of the $W$-boson self-energy in nuclear matter of
density $\rho$,
\begin{equation}
  W^{\mu\nu}(q) = -\frac{1}{\pi}\,\theta(\omega)\,
    \mathrm{Im}\,\Pi^{\mu\nu}_W(q;\rho),
  \label{eq:wfromself}
\end{equation}
with the self-energy and sign conventions of Refs.~\cite{gil1997,nieves2011}. Cutting a single particle--hole pair gives the quasi-elastic hyperon response of the
one-body current; cutting two particle--hole pairs, with the two-body operators of
Sec.~\ref{sec:twobody} at the vertices, gives the 2p2h response. The finite nucleus is
recovered in the local density approximation,
\begin{equation}
  W^{\mu\nu}_A(q) = \int d^3r\; W^{\mu\nu}\big(q;\rho(r)\big),
  \label{eq:lda}
\end{equation}
with $\rho(r)$ standing for the empirical proton and neutron densities of the target,
$\rho_p(r)$ and $\rho_n(r)$, which fix $k_F^p$ and $k_F^n$ locally.

\subsection{2p2h response}
\label{sec:2p2hresponse}

With the two-body current in momentum space written as
$j^\mu_{(2)}(p_1',p_2';p_1,p_2)$ for $N_1(p_1)N_2(p_2)\to Y(p_1')N'(p_2')$, the 2p2h tensor
per unit volume of matter with proton and neutron densities $\rho_p$ and $\rho_n$ is
\begin{widetext}
\begin{align}
  W^{\mu\nu}_{2p2h}(q;\rho) = {}&\sum_{N_1N_2;\,YN'} S_{N_1N_2;\,YN'}
    \int\frac{d^3p_1}{(2\pi)^3}\frac{d^3p_2}{(2\pi)^3} \nonumber\\
   &\times\;
    \frac{d^3p_1'}{(2\pi)^3}\frac{d^3p_2'}{(2\pi)^3}\,
    \frac{n_{N_1}(p_1)\,n_{N_2}(p_2)\,\big[1-n_{N'}(p_2')\big]}{16\,E_1E_2E_1'E_2'}
    \nonumber\\
   &\times\;
    (2\pi)^3\,\delta^4\!\big(q+p_1+p_2-p_1'-p_2'\big)\;
    \sum_{s_1s_2s_1's_2'}
      j^{\mu*}_{(2)}\,j^{\nu}_{(2)}\Big|_{N_1N_2\to YN'},
  \label{eq:2p2h}
\end{align}
\end{widetext}
where $E_i$ is the energy of baryon $i$ with its own mass,
$n_N(p)=\theta(k_F^N-|\mathbf{p}|)$ is the local-Fermi-gas occupation number with
$k_F^N=(3\pi^2\rho_N)^{1/3}$, the spinors are normalised covariantly, $\bar uu=2M$, like
the states of Eq.~\eqref{eq:hadronic}, and the index order is that of Eq.~\eqref{eq:hadronic},
$W^{\mu\nu}\propto\sum_f\langle f|J^\mu|i\rangle^*\langle f|J^\nu|i\rangle$, so that
$W^{\nu\mu}=W^{\mu\nu*}$. The spin sum runs over all four baryons, both spin states of every
occupied momentum being filled; the rate per pair given below, the unit of the factorised
benchmark, averages instead over the initial spins and momenta of one pair. The sum runs over the seven charge channels of Table~\ref{tab:channels}, each pair
once: for a $pn$ channel $N_1$ is the nucleon on line~1 of assignment $A$, and $j_{(2)}$ is
the coherent sum of Eq.~\eqref{eq:fullcurrent}. Integration variables label
ordered momenta, the amplitude is antisymmetrised in every pair of identical fermions,
and $S=2^{-n}$, with $n$ the number of identical pairs among the initial and final
particles, removes the multiple counting of each physical state that the ordering
introduces: $S=1$ for the three $pn$ channels, and $S=1/2$ for $pp\to\Lam p$, $\Sig^0p$,
$\Sig^+n$, $nn\to\Sig^-n$ and the non-strange mirror $pn\to nn$. For an
identical initial pair the antisymmetrised amplitude is
$\mathcal M_{\rm as}=\mathcal M(1,2)-\mathcal M(2,1)$, the two Wick contractions of the
normalised state $a^\dagger_1a^\dagger_2|0\rangle$, and
\begin{equation*}
  \frac{1}{2!}\sum_{s_1s_2}\!\int\! dp_1\,dp_2\,|\mathcal M_{\rm as}|^2
  = \sum_{s_1s_2}\!\int\! dp_1\,dp_2\,
    \Big[|\mathcal M(1,2)|^2-\mathrm{Re}\,\mathcal M(1,2)\mathcal M^*(2,1)\Big]:
\end{equation*}
per unordered pair, the unit in which the pair census of Sec.~\ref{sec:channels} counts
pairs, $Z(Z-1)/2$ for $pp$, the rate is $\langle|\mathcal M_{\rm as}|^2\rangle$, averaged
over the initial spins and the pair momenta, with no further factor: the
incoherent sum of the two orderings plus their exchange interference, the comparison
made in Table~\ref{tab:antisym}. The two final neutrons of the mirror are treated in the same
way: its two line assignments contain the contact and in-flight diagrams of each other
at exchanged final momenta, so the amplitude $X$ is the sum of the distinct diagrams and
the rate is $\tfrac12\sum|X(p_1',p_2')-X(p_2',p_1')|^2$ over ordered final momenta.

The hyperon is not Pauli blocked: the factor $[1-n(p_1')]$ of a knocked-out nucleon is
absent, which raises the 2p2h strength relative to the non-strange case at low
$\omega$. The hyperon carries its own in-medium potential $U_Y(\rho)$ in place of the
nucleon potential, which shifts the argument of the energy-conserving delta function
locally by $U_Y(\rho(r))-U_N(\rho(r))$; the shift depends on the species,
$U_\Lam(\rho_0)\approx-30\MeV$ being attractive~\cite{gal2016} and the $\Sig$-nucleus
potential repulsive~\cite{saha2004}, and is not evaluated here. Through the LDA integral
it ties the 2p2h response to the in-medium potential that hyperon-production
experiments seek to measure~\cite{companionPRD}. The elementary energy scale is
$M_Y-M_N\simeq177$--$258\MeV$, as for the quasi-elastic hyperon channel; at fixed
$|\mathbf q|$ the 2p2h threshold lies higher, moved by the recoil of the pair, Fermi
motion, Pauli blocking of $N'$, removal energies and $U_Y-U_N$, so the $\DS=1$ 2p2h
strength sits higher in $\omega$ than the quasi-elastic response, though the two overlap
(Sec.~\ref{sec:distributions}).

\subsection{Matching}
\label{sec:interference}
\label{sec:boundary}

The one-body current and the two-body classes can contribute to the same many-body
final states once correlations allow a one-body operator to populate multinucleon final
states, so the separation of ``QE'' and ``2p2h'' hyperon production is a statement
about the current operator. Their interference does not vanish by any symmetry, and the
non-strange argument for neglecting it, that the two responses peak in different
regions of $\omega$, is weaker here, since the $\DS=1$ 2p2h response starts at the
hyperon threshold. Four pieces must be distinguished:
\begin{enumerate}
\item the \textbf{irreducible two-body current}, with no intermediate state that can go
      on shell between the two vertices;
\item \textbf{reducible} amplitudes, one-body production followed by a real
      rescattering of the hyperon on a second nucleon, which are cascade steps and belong
      to the transport stage;
\item the \textbf{interference} between (1) and (2);
\item \textbf{double counting} at transport level, if (2) is included both in the
      current and in the cascade.
\end{enumerate}
The baryon-pole classes and the weak-first pole of the kaon-exchange family contain (1)
and (2) together: near the pole a hyperon produced by the one-body current rescatters on
the second nucleon, or is produced on the partner and transfers its strangeness by $\bar
K$ exchange. The off-shell prescription below removes the near-on-shell part (2) and
retains a virtual component (1); the identification is matching-scheme dependent. It
addresses (4); (3) is neither computed nor bounded.

\paragraph{Formation length.} The argument is the formation-zone one that transport
generators apply to produced hadrons~\cite{golan2012}, used here to define the current.
An intermediate baryon off shell by $\delta E = |p^2-M^2|/2M$ persists for a time
$\sim\!1/\delta E$ and propagates a distance $\ell \simeq v/\delta E$. If $\ell$ is much
smaller than the internucleon spacing $d$, the intermediate cannot reach the second
nucleon as a particle and the process is a two-body current; if $\ell\gtrsim d$ it is a
cascade step. Setting $\ell=d$ gives the boundary
\begin{equation}
  \Delta \;=\; |p^2-M^2|_{\rm crit} \;=\; \frac{2Mv\,\hbar c}{d},
  \label{eq:formation}
\end{equation}
with $v$ the dimensionless velocity, $d$ in fm and $\hbar c=0.19733\GeV\fm$. The
relevant $d$ runs from the short-range-correlated pair separation ($\simeq1.2\fm$) to
the mean nucleon spacing at the reduced densities of the nuclear surface
($\rho\simeq0.45\rho_0$, $\rho^{-1/3}\simeq2.4\fm$); at saturation the spacing is
$\rho_0^{-1/3}\simeq1.84\fm$. With $v=0.3$--$0.7$ and $M=M_N$, Eq.~\eqref{eq:formation}
gives $\Delta = 0.046$--$0.216\GeV^2$, centred on $0.10\GeV^2$. The same $\Delta$ is
applied to the hyperon-pole ordering, for which $M=M_\Lam$ or $M_\Sig$ would raise it by
a factor $1.2$--$1.3$.

\paragraph{Regulated principal value.} For a stable propagator,
$1/(x\pm i\epsilon) = \mathcal{P}(1/x) \mp i\pi\delta(x)$ motivates associating the
$\delta$-function term with on-shell propagation and the principal-value term with a
virtual contribution. We use the Lorentzian-regulated form $x/(x^2+\epsilon^2)$, which
tends to $\mathcal{P}(1/x)$ as $\epsilon\to0$ only as a distribution integrated against a
smooth test function; inside a squared amplitude with cuts it defines a model whose
$\epsilon$ does not drop out, and we treat $\epsilon$ as a matching scale on the same
footing as $\Delta$.

In the weak-first poles the near-on-shell intermediate is a hyperon, and the removed
part is one-body production followed by hyperon rescattering; in the absorption-first
pole it is a nucleon that absorbed the exchanged meson before the weak vertex, and the
removed part is one-body production on a nucleon whose momentum the partner has
changed, which a quasi-elastic calculation with a realistic spectral function contains.
Both are removed by the same rule. Table~\ref{tab:dctable} sets each class beside the
treatment of its non-strange analogue in the Valencia~\cite{nieves2011} and
Torino--Seville~\cite{amaro2017} models.

\begin{table}[t]
\caption{\label{tab:dctable}Treatment of each class of the $\DS=1$ current, and of its
non-strange analogue (in parentheses) in two microscopic 2p2h models. ``Regulated'':
the near-on-shell part removed by the cut or suppressed by the principal value of
Sec.~\ref{sec:interference}. NP: nucleon pole; $\Delta$P: $\Delta$ pole.}
\begin{ruledtabular}
\begin{tabular}{llll}
\parbox[t]{0.30\textwidth}{\raggedright class (non-strange analogue)\strut} &
  \parbox[t]{0.17\textwidth}{\raggedright here\strut} &
  \parbox[t]{0.26\textwidth}{\raggedright Valencia~\cite{nieves2011}\strut} &
  \parbox[t]{0.17\textwidth}{\raggedright Torino--Seville~\cite{amaro2017}\strut} \\
\hline
\parbox[t]{0.30\textwidth}{\raggedright (a) contact (Kroll--Ruderman seagull)\strut} &
  \parbox[t]{0.17\textwidth}{\raggedright kept\strut} &
  \parbox[t]{0.26\textwidth}{\raggedright kept\strut} &
  \parbox[t]{0.17\textwidth}{\raggedright kept\strut} \\
\parbox[t]{0.30\textwidth}{\raggedright (b) meson-in-flight (pion-in-flight)\strut} &
  \parbox[t]{0.17\textwidth}{\raggedright kept\strut} &
  \parbox[t]{0.26\textwidth}{\raggedright kept\strut} &
  \parbox[t]{0.17\textwidth}{\raggedright kept\strut} \\
\parbox[t]{0.30\textwidth}{\raggedright (c1) weak-first pole, hyperon intermediate (NP, $W$ first)\strut} &
  \parbox[t]{0.17\textwidth}{\raggedright kept, regulated\strut} &
  \parbox[t]{0.26\textwidth}{\raggedright NP--NP subtracted as part of QE; NP interference terms kept, small width on the nucleon\strut} &
  \parbox[t]{0.17\textwidth}{\raggedright omitted (spectral function, FSI)\strut} \\
\parbox[t]{0.30\textwidth}{\raggedright (c2) absorption-first pole, nucleon intermediate (crossed NP, meson first)\strut} &
  \parbox[t]{0.17\textwidth}{\raggedright kept, regulated\strut} &
  \parbox[t]{0.26\textwidth}{\raggedright crossed NP kept (square and interference)\strut} &
  \parbox[t]{0.17\textwidth}{\raggedright omitted\strut} \\
\parbox[t]{0.30\textwidth}{\raggedright (d) decuplet poles: direct $\Sst$, crossed $\Delta$, $\Sst$ on the partner ($\Delta$P, crossed $\Delta$P)\strut} &
  \parbox[t]{0.17\textwidth}{\raggedright kept, width only, not cut\strut} &
  \parbox[t]{0.26\textwidth}{\raggedright kept; $\Delta$P--$\Delta$P subtracted against $\Gamma_\Delta^{\rm Pauli}$ and $C_{A2}$\strut} &
  \parbox[t]{0.17\textwidth}{\raggedright kept\strut} \\
\parbox[t]{0.30\textwidth}{\raggedright (e) kaon exchange, hyperon pole on the partner (NP with the $W$ on the partner)\strut} &
  \parbox[t]{0.17\textwidth}{\raggedright kept, regulated\strut} &
  \parbox[t]{0.26\textwidth}{\raggedright as (c1)\strut} &
  \parbox[t]{0.17\textwidth}{\raggedright omitted\strut} \\
\parbox[t]{0.30\textwidth}{\raggedright (e) line-2 kaon seagull (contact with the $W$ on the partner)\strut} &
  \parbox[t]{0.17\textwidth}{\raggedright kept\strut} &
  \parbox[t]{0.26\textwidth}{\raggedright kept\strut} &
  \parbox[t]{0.17\textwidth}{\raggedright kept\strut} \\
\end{tabular}
\end{ruledtabular}
\end{table}

\subsection{Correlations}
\label{sec:srcdef}

In the reference evaluation every exchanged meson carries the bare propagator of
Eq.~\eqref{eq:line2}. Short-range correlations are a property of the $NN$ pair, so a
consistent treatment has to address every exchanged meson, although not necessarily
with the same function for each. Two correlated treatments are carried as variants. The
first is the Landau--Migdal contact of the spin-longitudinal pion channel in the form of
Oset, Toki and Weise and of Nieves \emph{et al.}~\cite{oset1982,nieves2011},
\begin{equation}
  V_L(k) = \frac{f_{\pi NN}^2}{m_\pi^2}\Big[F_\pi^2(k^2)\,
    \frac{|\mathbf k|^2}{k^2-m_\pi^2}+g'\Big],\qquad g'=0.7 ,
  \label{eq:vl}
\end{equation}
with $k^2=(k^0)^2-|\mathbf k|^2$ and $\mathbf k$ the three-momentum in the rest
frame of the nucleus. On the pseudovector line, with $\sla k\gamma_5$ at both vertices
and the single monopole of Eq.~\eqref{eq:line2} in place of $F_\pi^2$,
it becomes $\widetilde D_\pi\to F_\pi/(k^2-m_\pi^2)+g'/|\mathbf k|^2$, the contact carrying
no form factor. For spacelike exchange the contact is repulsive, and at the momenta a
correlated pair supplies, $|\mathbf k|\simeq0.4$--$0.8\GeV$, it cancels most of the pion
pole. It acts on pions only. The second is one Oset--Toki--Weise correlation function,
$g(r)=1-j_0(q_cr)$ with $q_c=0.78\GeV$, applied to every meson exchanged between the
nucleons ($\pi$, $\eta$ and $K$). In momentum space it subtracts the exchange averaged
over the shell $|\mathbf k'-\mathbf k|=q_c$. In the approximation of
Ref.~\cite{oset1982}, which averages the numerator exactly and evaluates the propagator at
the mean squared momentum of the shell, $|\mathbf k|^2+q_c^2$, the subtraction is the same
exchange at $|\mathbf k|^2\to|\mathbf k|^2+q_c^2$ plus a spin--spin contact whose
longitudinal part is carried on the $\sla k\gamma_5$ line,
\begin{equation}
  \widetilde D_\phi^{\rm corr}(k) = \frac{F_\phi(k^2)}{k^2-m_\phi^2}
   - \frac{F_\phi(k^2-q_c^2)}{k^2-q_c^2-m_\phi^2}
   - \frac{q_c^2}{3\,|\mathbf k|^2}\,\frac{F_\phi(k^2-q_c^2)}{k^2-q_c^2-m_\phi^2};
  \label{eq:src}
\end{equation}
its transverse remainder has no counterpart in this Dirac structure and is dropped.
Both are medium dressings,
switched off in the symmetry tests (Sec.~\ref{sec:ward}).

\subsection{Pair approximation}
\label{sec:factorised}

Equation~\eqref{eq:2p2h} delivers the lepton kinematics, not the hadronic final state a
detector simulation requires. Generators use a factorised prescription --- sample a
correlated pair from a short-range-correlation momentum distribution, apply the
two-body current to it, distribute the three-body final state according to phase space
--- which reproduces an inclusive response only if its pair distribution and current
weights are matched to the many-body calculation. The evaluation here takes the pair at
zero total momentum, $\mathbf{p}_2 = -\mathbf{p}_1$, with $|\mathbf p_1|=0.25\GeV$, and
averages the current over the direction of $\mathbf p_1$, taken isotropic. This
reduces Eq.~\eqref{eq:2p2h} to a three-body phase-space integral weighted by
$|j_{(2)}|^2$ per correlated pair.

\section{Validation}
\label{sec:results}

The class operators of Sec.~\ref{sec:classes} are evaluated by explicit Dirac spin sums
over a three-body phase space, in the factorised approximation of
Sec.~\ref{sec:factorised}, with classes (a)--(e) all included.

\subsection{Scope}
\label{sec:layers}

The lowest-order chiral content is the operator topology of Sec.~\ref{sec:classes}, the
flavour-route coefficients of Sec.~\ref{sec:isospin}, the Noether vector construction and
the Goldberger--Treiman completion; masses, propagators, form factors, widths, decuplet
couplings, matching and short-range correlations are imported, so the current is a
tree-level chiral model, not complete at a definite chiral
order~\cite{krebs2020,baroni2016}. The nuclear-matter tensor of
Eq.~\eqref{eq:2p2h} is formulated but not evaluated; the evaluated quantities are ratios
between classes, hyperon channels and variants of the current on the factorised
correlated pair.

\subsection{Inputs}
\label{sec:inputs}

\begin{table}[t]
\caption{\label{tab:inputs}Inputs of the reference evaluation and the variations run.
The reference carries no short-range correlation; $g'$ and $q_c$ belong to the two
correlated variants of Sec.~\ref{sec:srcdef}, and $\epsilon$ to the alternative
matching scheme of Sec.~\ref{sec:interference}. The pair direction is isotropic.}
\begin{ruledtabular}
\begin{tabular}{lll}
input & reference & variations \\
\hline
$D$, $F$ & $0.804$, $0.467$ & $\pm0.008$ each \\
$f_\pi=f_K$ & $92.4\MeV$ & --- \\
$f_+(0)$, slope $\lambda_+$ & $0.970$, $0.0297$ & --- \\
dipole masses $M_V$, $M_A$ & $0.84$, $1.0\GeV$ & $M_A=0.9$, $1.2\GeV$ \\
monopole cutoff $\Lambda_\pi=\Lambda_K$ & $1.3\GeV$ & $1.0$, $1.5\GeV$ \\
$C_{3,4,5}^V(0)$; $C_5^A(0)$, $M_A^\Delta$ ($n\!\to\!\Delta^+$) & $2.13$, $-1.51$, $0.48$; $1.2$, $0.95\GeV$ & --- \\
$\Gamma_{\Sig^*}$; $g_{\Sig^*\Lam\pi}$, $g_{\Sig^*\Sig\pi}$ & $36\MeV$; $1.27$, $0.95$--$1.09$ & $\Gamma_{\Sig^*}=61.5$, $100\MeV$ \\
$\Gamma_\Delta$ & $117\MeV$ & $200\MeV$ \\
Landau--Migdal $g'$ (pions only) & none & $0.7$ (scan $0$--$0.9$) \\
correlation $q_c$ (every meson) & none & $0.78\GeV$ ($0.60$, $1.00\GeV$) \\
formation-length cut $\Delta$ & $0.10\GeV^2$ & $0.046$, $0.216\GeV^2$ \\
principal value $\epsilon$ & --- & $0.03\GeV^2$ ($0.01$--$0.10\GeV^2$) \\
pair momenta & $\mathbf p_2=-\mathbf p_1$, $|\mathbf p_1|=0.25\GeV$ & --- \\
momentum transfer & $|\mathbf q|<1.2\GeV$ & $1.1$, $1.3\GeV$, none \\
$E_{\nubar}$ & $0.6$--$5\GeV$ & $10\GeV$ (diagnostic) \\
pair weights, $N_{pn}/N_{pp}$ & $18$: $87.8\%$ $pn$, $4.9\%$ $pp$, $7.4\%$ $nn$ & $12$--$24$ \\
\end{tabular}
\end{ruledtabular}
\end{table}

The reference evaluation uses the inputs of Table~\ref{tab:inputs} inside
$|\mathbf q|<1.2\GeV$, where non-strange 2p2h calculations carry most of their strength;
whether the strange response is exhausted below the same cutoff is not shown. Raising the
cutoff from $1.1$ to $1.3\GeV$~\cite{gran2013} raises the coherent $\DS=1$ rate by $3\%$
at $1\GeV$ and $7$--$13\%$ at $1.5$--$5\GeV$ and moves $\sigma_\Lam/\sigma_{\Sig^0}$ by
at most $0.9\%$; without the restriction the pair strength rises by $46\%$ at $3\GeV$ and
by a factor $1.84$ at $5\GeV$, and $\sigma_\Lam/\sigma_{\Sig^0}$ by $18\%$ and $35\%$.
Above $\sim\!5\GeV$ the free pair opens phase space the nucleus does not have, so the
$10\GeV$ rows of the tables are diagnostics of the surrogate and results are quoted
for $0.6$--$5\GeV$.

\subsection{Benchmarks}
\label{sec:validation}

Applied to the one-body current, the code path --- spinors, leptonic contraction,
prefactor, phase space and Eq.~\eqref{eq:xsec} --- gives
$\sigma(\nubar p\to\mu^+n)=3.61\times10^{-39}\cm^2$ at $1\GeV$ against
$3.66\times10^{-39}\cm^2$ from the Llewellyn Smith formula with the same dipole form
factors, and differs from the quasi-elastic hyperon channel of the generator of
Ref.~\cite{companionPRD} by up to $2\%$ (Table~\ref{tab:validation}); both differences
are the fluctuation of one seed of $3\times10^4$ samples. The checks exercise the vertex,
phase space and cross-section assembly, not the route structure, off-shell prescription
or pair model. The $\DS=0$ control also fixes the sign of the vector--axial
interference, which no $\DS=1$ check sees and which enters the ratio of
Ref.~\cite{Nowak:2026pheno}.

\begin{table}[b]
\caption{\label{tab:validation}One-body quasi-elastic hyperon cross section per
nucleon [cm$^2$], argon-weighted over protons and neutrons with the struck nucleon at
rest, computed with the machinery of this section ($3\times10^4$ Monte Carlo samples
per point, one seed for every channel) against the quasi-elastic channel of the generator of
Ref.~\cite{companionPRD} with the same definition ($4\times10^5$ samples).}
\begin{ruledtabular}
\begin{tabular}{lrrr}
$E_{\nubar}$ [GeV] & this work & Ref.~\cite{companionPRD} & ratio \\
\hline
1.0  & $9.66\times10^{-41}$ & $9.78\times10^{-41}$ & 0.988 \\
2.0  & $1.57\times10^{-40}$ & $1.60\times10^{-40}$ & 0.983 \\
3.0  & $1.83\times10^{-40}$ & $1.87\times10^{-40}$ & 0.981 \\
5.0  & $2.07\times10^{-40}$ & $2.10\times10^{-40}$ & 0.983 \\
10.0 & $2.29\times10^{-40}$ & $2.30\times10^{-40}$ & 0.996 \\
\end{tabular}
\end{ruledtabular}
\end{table}

\subsection{Symmetry tests}
\label{sec:ward}

\paragraph{One- and two-body identities.} The reduction of Eq.~\eqref{eq:ward1b} with
Eq.~\eqref{eq:g3pcac} holds in the implementation to $4\times10^{-14}$ for all three
hyperon channels, a test of $g_3$ that a cross section, in which the induced
pseudoscalar enters suppressed by $m_\ell^2$, cannot make. In the chiral and
SU(3)-degenerate limit both divergences of a complete two-body current vanish
identically; the one-body control inside three-body phase space is contracted with its
own momentum transfer $p_1'-p_1=q+k_\pi$ and returns zero, the two-body current with
$q$, which the $W$ injects and the exchanged meson redistributes.

\paragraph{Diagnostics.} The chiral-limit conservation residual and the physical
longitudinal fraction share one functional form at different masses,
\begin{equation}
  \delta_A^{\chi}\;(\text{chiral limit}),\ \ L_A\;(\text{physical})
  \;=\;
  \frac{|q\!\cdot\!A|}{\sum_\mu |q^\mu||A^\mu|+\epsilon},
  \label{eq:deltadef}
\end{equation}
with $A=A_{\rm comp}$: $\delta_A^\chi$ tests conservation when explicit breaking is
removed, and $L_A$ measures how longitudinal the physical amplitude is, a nonzero $L_A$
being the kaon-pole divergence PCAC requires. The mismatch of the longitudinal completion
at physical masses is
\begin{equation}
  \delta_{\rm long}
  \;=\;
  \frac{\big|q\!\cdot\!A_{\rm comp}-(m_s+m_u)P_{su}\big|}
       {|q\!\cdot\!A_{\rm comp}|+|(m_s+m_u)P_{su}|+\epsilon},
  \label{eq:dlong}
\end{equation}
with $(m_s+m_u)P_{su}$ assembled chain by chain over the same routes and prescriptions
as $A^\mu_{\rm comp}$; $\delta_V$ uses
Eq.~\eqref{eq:deltadef} with $V$, and $\epsilon$ only prevents $0/0$. Each is evaluated
per phase-space point and spin configuration at $E=2\GeV$. With exact propagators and no
dressing the raw coherent divergences in the symmetric limit are
\begin{equation}
  \big\langle|q\!\cdot\!V|\big\rangle = 2.4\times10^{-7},
  \qquad
  \big\langle|q\!\cdot\!A|\big\rangle = 1.5\times10^{-5},
  \label{eq:wardresult}
\end{equation}
against a current scale $\langle|q||V|\rangle = 3.8\times10^{3}$ in the same units, the
axial entry being the $m_K^2$ retained by the scaled meson masses. Pointwise the bare
current closes at machine precision for both line assignments and the four like-pair
channels, medians $\delta_V^\chi=2\times10^{-10}$, $\delta_A^\chi=5\times10^{-9}$ and
$\delta_{\rm long}=1.3\times10^{-15}$ (99th percentiles in Table~\ref{tab:wardladder}),
and the physical amplitude is $7\%$ longitudinal, $L_A=0.069$ against $0.17$ for the
one-body control.

\paragraph{Axial contact operator.} Switched on alone, the axial contact operator of
Sec.~\ref{sec:constraints} breaks chiral-limit conservation: at the Lagrangian strength
$g=-1$ the median $\delta_A^\chi$ rises from $5\times10^{-9}$ to $1.1\times10^{-1}$, and
to $3$--$11\times10^{-2}$ at $|g|=0.25$--$1$. With its kaon-pole partner at the same
coefficient the median returns to the bare level ($4\times10^{-9}$ at $2\GeV$;
$2\times10^{-8}$ and $2\times10^{-9}$ at $1$ and $3\GeV$, as for the bare current), and
the vector residual is bit-identical. At physical masses the pair leaves
$\delta_{\rm long}$ a median of $1.5$--$6\times10^{-2}$ over $1$--$3\GeV$, the
SU(3)-breaking term of Sec.~\ref{sec:constraints}; with that term on the pseudoscalar
side it returns to below $4\times10^{-15}$.

\paragraph{Matching prescriptions.} Before dressing, the only cost to the identities is
the off-shell matching of Sec.~\ref{sec:boundary} (Table~\ref{tab:wardladder}). It falls
on the vector sector alone and is localised: the cut zeroes a chain at $4.8\%$ of the
points, where $\delta_V$ reaches $0.92$, and leaves the identity intact elsewhere; the
principal value costs an order of magnitude less. No longitudinal-restoration
counterterm is applied.

\begin{table}[t]
\caption{Symmetry cost of the matching prescriptions for line assignment $A$ of
Eq.~\eqref{eq:fullcurrent}. $\delta_V^\chi$ (p99, SU(3)$+$chiral limit) and
$\delta_{\rm long}$ (p99, physical masses) per stage.}
\label{tab:wardladder}
\begin{ruledtabular}
\begin{tabular}{lcc}
stage & $\delta_V^\chi$ (p99) & $\delta_{\rm long}$ (p99) \\
\hline
bare Feynman current            & $3\times10^{-9}$   & $3\times10^{-14}$ \\
cut, $\Delta=0.046\GeV^2$       & $4.2\times10^{-2}$ & $3\times10^{-14}$ \\
cut, $\Delta=0.10\GeV^2$        & $1.0\times10^{-1}$ & $3\times10^{-14}$ \\
cut, $\Delta=0.216\GeV^2$       & $4.6\times10^{-1}$ & $3\times10^{-14}$ \\
PV, $\epsilon=0.01\GeV^2$       & $1.2\times10^{-3}$ & $3\times10^{-14}$ \\
PV, $\epsilon=0.03\GeV^2$       & $9.5\times10^{-3}$ & $3\times10^{-14}$ \\
PV, $\epsilon=0.10\GeV^2$       & $7.4\times10^{-2}$ & $3\times10^{-14}$ \\
\end{tabular}
\end{ruledtabular}
\end{table}

\paragraph{Dressings.} The reference form factors (Table~\ref{tab:dressings}), applied
vertex by vertex, break the vector identity~\cite{grossriska1987} at a median residual
of $0.27$, because they weight the members of a Ward-related set by different powers of
$Q^2$; the axial construction is untouched, and all dressings are off in the symmetry
tests above. Conservation for a dressed current needs a common factor on a whole
Ward-related set or an explicit longitudinal restoring
term~\cite{grossriska1987,haberzettl1998,haberzettl2021}. One dipole $G_V(Q^2)$ on every
vector insertion, with each monopole-dressed exchange written as a difference of
propagators, restores the identity here, to a median $\delta_V^\chi=2.1\times10^{-10}$,
and at physical masses changes the channel-summed rate by under $3\%$ and
$\sigma_\Lam/\sigma_{\Sig^0}$ by at most $4.3\%$
(Table~\ref{tab:dressscheme}); the results below include the central cut, which breaks
the identity on its own. Without any form factor the pair strength is $2.5$--$7.8$ times
the reference and the composition moves by up to $15\%$: the rate depends on the
ultraviolet regulators, the composition far less.

\begin{table}[t]
\caption{\label{tab:dressscheme}The Ward-compatible and undressed currents against the
reference (per-vertex) dressing, central cut, no short-range correlation: seed-paired
change of $\sigma_\Lam/\sigma_{\Sig^0}$ and channel-summed coherent rate relative to the
reference. Five replicas of $2\times10^4$ samples, standard error of the mean on the last
digits in parentheses where it is at least half a unit.}
\begin{ruledtabular}
\begin{tabular}{lcccc}
 & \multicolumn{2}{c}{Ward-compatible} & \multicolumn{2}{c}{undressed} \\
\cline{2-3}\cline{4-5}
$E_{\nubar}$ [GeV] & $\Delta(\sigma_\Lam/\sigma_{\Sig^0})$ & $\sigma/\sigma_{\rm ref}$
 & $\Delta(\sigma_\Lam/\sigma_{\Sig^0})$ & $\sigma/\sigma_{\rm ref}$ \\
\hline
0.6 & $-4.3(1)\%$  & 0.974(1) & $-5.3(1)\%$  & 2.499(3) \\
1.0 & $-4.1\%$     & 0.972    & $-15.2(4)\%$ & 5.43(3) \\
2.0 & $-0.6\%$     & 0.994    & $-10.8(6)\%$ & 7.21(7) \\
3.0 & $+0.1(1)\%$  & 0.999(1) & $-9.0(15)\%$ & 7.79(15) \\
5.0 & $+0.2(1)\%$  & 1.000    & $-9.9(42)\%$ & 7.83(7) \\
\end{tabular}
\end{ruledtabular}
\end{table}

\begin{table}[t]
\caption{The phenomenological dressings: where each enters, and what each preserves.
The constructed longitudinal relation in the last column holds within the truncation of
Sec.~\ref{sec:constraints}.}
\label{tab:dressings}
\begin{ruledtabular}
\begin{tabular}{lccc}
dressing & enters at & $q\!\cdot\!V$ (SU(3) limit) & \begin{tabular}[b]{@{}c@{}}$q\!\cdot\!A_{\rm comp}=$\\$(m_s+m_u)P_{su}$\end{tabular} \\
\hline
weak dipoles $F_V,F_A$       & pole vertices      & broken (median $0.27$) & preserved \\
strong monopoles $F_\pi,F_K$ & each strong vertex & broken (with above)    & preserved \\
$K_{\ell3}$ $f_+(t)$         & in-flight weak vertex & broken (with above) & preserved \\
Landau--Migdal $g'$ (variant) & exchanged $\pi$    & broken (in flight)     & preserved \\
correlation $q_c$ (variant)  & every exchange     & broken (in flight)     & preserved \\
resonance width $\Gamma_R$   & RS propagator      & preserved (transverse) & preserved \\
$|\mathbf q|<1.2\GeV$        & phase space        & preserved              & preserved \\
formation-length cut $\Delta$       & pole chains        & broken ($10^{-1}$ p99) & preserved \\
principal value $\epsilon$   & pole chains        & broken ($10^{-2}$ p99) & preserved \\
\hline
Ward-compatible (common) & all vector insertions & preserved ($2.1\times10^{-10}$) & preserved \\
\end{tabular}
\end{ruledtabular}
\end{table}

\paragraph{Constraints from the identities.} The vector identity fixes the relative
flavour coefficients of the octet classes, the momentum routing of the pole classes
($p_1+k_\pi$, not $p_1-k_\pi$, for the absorption-first intermediate) and the
longitudinal operator structure. It does not constrain the transverse off-shell vertex,
the decuplet couplings, the relative weight and sign of the two line assignments (each
separately conserved), the overall normalisation, or anything that multiplies a
Ward-related set uniformly.

\section{Results}
\label{sec:numresults}

\subsection{Class hierarchy}
\label{sec:hierarchy}
\label{sec:decomposition}

\begin{figure*}[t]
\centering
\includegraphics[width=0.92\textwidth]{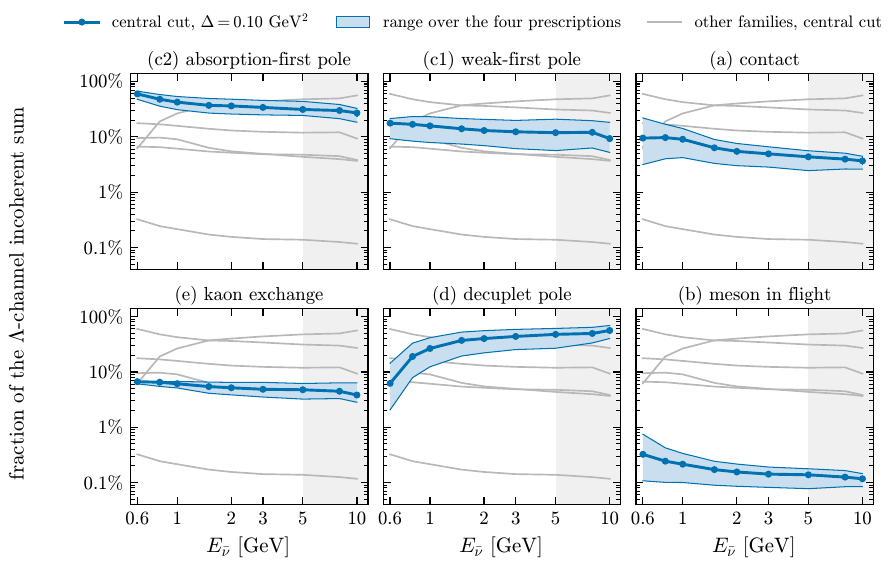}
\caption{\label{fig:hierarchy}Insertion families of the $\Lam$-channel current as
fractions of the incoherent family sum, against antineutrino energy, with no short-range
correlation and $|\mathbf q|<1.2\GeV$. Points: the central cut, $\Delta=0.10\GeV^2$,
mean of five replicas of $2\times10^4$ samples with the standard error of the mean.
Band: range of the replica means over the four matching prescriptions of
Table~\ref{tab:hierpresc}. Grey lines: the other five families at the central cut.
Shaded: $E_{\nubar}>5\GeV$ (Sec.~\ref{sec:inputs}).}
\end{figure*}

Table~\ref{tab:classes} gives each insertion family as a fraction of the incoherent sum
for the $\Lam$ channel, and Fig.~\ref{fig:hierarchy} shows the families against energy.
Every class is evaluated on the same kinematics, phase space and pair treatment, so the
ratios are free of the overall normalisation.

The meson-in-flight class, a leading long-range term of the non-strange current where
the $\Delta$-pole mechanism and its interference carry most of the
response~\cite{nieves2011,gran2013}, is $0.1$--$0.3\%$ of the incoherent sum in the
$\Lam$ channel from $0.6$ to $5\GeV$ and below $0.06\%$ in the $\Sig$ channels, where
the $KN\Sig$ coupling is the small $D-F$ combination, as the range argument of
Sec.~\ref{sec:apriori} anticipates. The share of the two pole orderings falls from
$77\%$ at $0.6\GeV$ to $43$--$58\%$ at $1$--$5\GeV$ while the decuplet rises from $6\%$
to $26$--$48\%$; the seagull carries $4$--$10\%$ and kaon exchange $5$--$7\%$. In the
$\Sig$ channels the poles carry $56$--$93\%$, the seagulls are $D-F$-suppressed and kaon
exchange is about $1\%$.

\begin{table}[b]
\caption{\label{tab:classes}Insertion families of the completed current as fractions of
the incoherent sum, $\Lam$ channel, at the default formation-length cut $\Delta=0.10\GeV^2$,
with the momentum-transfer restriction $|\mathbf q|<1.2\GeV$ and no short-range
correlation on any exchanged meson. Means of the five replicas of $2\times10^4$ samples
of Fig.~\ref{fig:hierarchy}, with the standard error of the mean on the last digits in
parentheses where it is at least half a unit.
``pole (w)'' and ``pole (a)'' are the line-1 weak-first and absorption-first orderings,
and ``poles'' their sum;
``kaon exch.''\ is the line-2 family of Eq.~\eqref{eq:kexch}, and ``decuplet'' collects
the direct and crossed decuplet poles, their $\eta$ routes and the line-2 $\Sst$. Every
exchange, charge route and line assignment is folded into its family.}
\begin{ruledtabular}
\begin{tabular}{lrrrrrrr}
$E_{\nubar}$ [GeV] & seagull & in-flight & pole (w) & pole (a) & decuplet & kaon exch. & poles \\
\hline
0.6  & 0.095(1)  & 0.0032    & 0.177(1)  & 0.596(2)  & 0.062(1)  & 0.067     & 0.773(1) \\
1.0  & 0.090     & 0.0022    & 0.158(1)  & 0.424(2)  & 0.265(2)  & 0.061     & 0.582(2) \\
2.0  & 0.055     & 0.0016    & 0.130(2)  & 0.362(2)  & 0.400(4)  & 0.052(1)  & 0.492(3) \\
3.0  & 0.049(1)  & 0.0014    & 0.123(3)  & 0.340(8)  & 0.438(11) & 0.049(2)  & 0.463(10) \\
5.0  & 0.044(2)  & 0.0014    & 0.119(3)  & 0.312(18) & 0.477(21) & 0.048(2)  & 0.431(19) \\
10.0 & 0.037(6)  & 0.0012(1) & 0.093(12) & 0.270(40) & 0.560(32) & 0.038(5)  & 0.363(34) \\
\end{tabular}
\end{ruledtabular}
\end{table}

\begin{table}[t]
\caption{\label{tab:leaveoneout}Incoherent fractions and coherent leave-one-out weights
of the $\Lam$-channel families at the central prescription, with common random numbers
($8\times10^3$ samples, one seed; incoherent fractions from the same run).}
\begin{ruledtabular}
\begin{tabular}{lcccc}
 & \multicolumn{2}{c}{incoherent fraction} & \multicolumn{2}{c}{coherent leave-one-out} \\
family & $1\GeV$ & $3\GeV$ & $1\GeV$ & $3\GeV$ \\
\hline
contact (seagull)        & $9.3\%$  & $4.9\%$  & $+0.3\%$  & $-2.2\%$ \\
meson-in-flight          & $0.22\%$ & $0.14\%$ & $0.0\%$   & $+0.1\%$ \\
pole (weak-first)        & $16.0\%$ & $11.9\%$ & $+12.9\%$ & $+7.0\%$ \\
pole (absorption-first)  & $44.0\%$ & $36.1\%$ & $+34.0\%$ & $+24.2\%$ \\
decuplet                 & $24.4\%$ & $42.1\%$ & $+36.3\%$ & $+56.0\%$ \\
kaon exchange            & $6.2\%$  & $4.9\%$  & $-0.5\%$  & $-1.2\%$ \\
\hline
pole sector, both orderings removed together & $59.9\%$ & $47.9\%$ & $+53.6\%$ & $+40.3\%$ \\
\end{tabular}
\end{ruledtabular}
\end{table}

\paragraph{Prescription dependence.} The current-to-transport separation regulates the
octet-baryon propagators only, so the seagull, in-flight and decuplet amplitudes and the
line-2 kaon seagull are bit-identical under every prescription of
Table~\ref{tab:hierpresc}, and their shares move only with the total. Under every
prescription and at every energy the in-flight class stays below $1\%$ and kaon exchange
at $3$--$7\%$.

\begin{table}[t]
\caption{Insertion families of the $\Lam$ channel as fractions of the incoherent sum,
under each current-to-transport prescription, at two energies, in the reference
evaluation (no short-range correlation, $|\mathbf q|<1.2\GeV$). ``poles'' is the sum
of the two orderings, with the Monte Carlo error on its last digit in parentheses,
from five independent replicas of $2\times10^4$ samples per point (independent in the
sense of Appendix~\ref{sec:repro}: the same calculation with every generator reseeded).}
\label{tab:hierpresc}
\begin{ruledtabular}
\begin{tabular}{lcccccccc}
prescription & $E$ [GeV] & seagull & in-fl. & pole (w) & pole (a) & dec. & K exch. & poles \\
\hline
cut, $0.046$ & 1 & 0.049 & 0.0012 & 0.217 & 0.522 & 0.145 & 0.065 & 0.739(3) \\
             & 3 & 0.032 & 0.0009 & 0.182 & 0.442 & 0.281 & 0.063 & 0.624(11) \\
cut, $0.10$  & 1 & 0.090 & 0.0022 & 0.158 & 0.424 & 0.265 & 0.061 & 0.582(2) \\
             & 3 & 0.049 & 0.0014 & 0.123 & 0.340 & 0.438 & 0.049 & 0.463(10) \\
cut, $0.216$ & 1 & 0.141 & 0.0034 & 0.079 & 0.310 & 0.415 & 0.051 & 0.389(1) \\
             & 3 & 0.066 & 0.0019 & 0.061 & 0.249 & 0.586 & 0.035 & 0.311(5) \\
PV, $0.03$   & 1 & 0.042 & 0.0010 & 0.231 & 0.535 & 0.124 & 0.067 & 0.766(2) \\
             & 3 & 0.028 & 0.0008 & 0.198 & 0.455 & 0.253 & 0.065 & 0.653(12) \\
\end{tabular}
\end{ruledtabular}
\end{table}

\paragraph{Pole dominance.} Pole dominance is a property of the low-energy region and of
the narrow matching scales. The poles carry more than all other families together at
$0.6\GeV$ under every prescription, up to $1.5\GeV$ at the central cut, and up to $5\GeV$
under the narrow cut and the principal value, which retain more of the off-shell pole
strength; at the central cut the decuplet equals them at $3\GeV$ and exceeds them above.
Over $0.6$--$5\GeV$ and the four prescriptions the pole share runs from $30\%$ to $88\%$.

\paragraph{Coherent leave-one-out weights.} A fraction $\sigma_i/\sum_j\sigma_j$ of
separately evaluated families is not an observable decomposition when the amplitudes
interfere. The coherent leave-one-out weight,
$1-\sigma(\text{without }i)/\sigma(\text{all})$, measures what the rate does when a
family is removed from the amplitude; only the coherent total is basis invariant.
Table~\ref{tab:leaveoneout} gives both measures. The seagull's coherent weight is
$+0.3\%$ at $1\GeV$, against an incoherent fraction of $9.3\%$, and $-2.2\%$ at $3\GeV$:
its interference with the rest of the current cancels its own square. The in-flight
family is $0.1\%$ of the coherent rate. Single weights cannot be added, interference
being counted in every entry that participates in it: removing both pole orderings
removes $54\%$ of the rate at $1\GeV$ and $40\%$ at $3\GeV$, against their incoherent
$60\%$ and $48\%$.

\paragraph{Interference among the classes.} The six families of Table~\ref{tab:classes}
interfere destructively: summed over the $\Lam$, $\Sig^0$ and $\Sig^-$ channels, the
coherent rate is $14\%$ below the incoherent sum at $0.6\GeV$ and $21$--$22\%$ below it
over $1$--$5\GeV$ (Table~\ref{tab:interference}). Every cross section and cross-section
ratio below is computed from the coherent sum.

\begin{table}[t]
\caption{\label{tab:interference}Coherent over incoherent $\DS=1$ rate, summed over the
$\Lam$, $\Sig^0$ and $\Sig^-$ channels, at the central cut. Means of three independent
replicas of $2.5\times10^4$ samples, with the Monte Carlo error on the last digits in
parentheses.}
\begin{ruledtabular}
\begin{tabular}{lcccccc}
$E_{\nubar}$ [GeV] & 0.6 & 1.0 & 2.0 & 3.0 & 5.0 & 10.0 \\
\hline
coherent/incoherent & $0.858(3)$ & $0.786(2)$ & $0.778(3)$ & $0.779(4)$ & $0.779(7)$ & $0.790(2)$ \\
\end{tabular}
\end{ruledtabular}
\end{table}

\paragraph{Line assignments.} The two line assignments of Eq.~\eqref{eq:fullcurrent}
are each separately conserved, so the symmetry tests do not fix their relative weight.
Table~\ref{tab:assign} separates them: $\sigma_A$ and $\sigma_B$ are the coherent rates
with one assignment alone, $B$ being the one that carries the charged-pion routes, and
$I=\sigma_{A+B}-\sigma_A-\sigma_B$ their interference. Assignment $B$ carries
$89$--$91\%$ of the coherent channel-summed rate and assignment $A$ $24$--$27\%$ at every
energy from $0.6$ to $5\GeV$, and the interference is destructive in every channel and
at every energy, $-15$ to $-17\%$ of the channel-summed rate. In assignment $B$ the
final hyperon comes from the nucleon the naive charge rule does not assign it to, the
configuration an isospin-formalism treatment counts as the exchange term. Above $1\GeV$
assignment $B$ alone gives $\sigma_\Lam/\sigma_{\Sig^0}$ within $6\%$ of the complete
current, and assignment $A$ alone $1.4$--$1.6$ times it: the composition is set by
assignment $B$, and the interference, most destructive in the $\Lam$ channel, makes the
complete current less $\Lam$-rich than either assignment alone. In the $\Sig^-$ channel
assignment $B$ has no seagull and no kaon-exchange term.

\begin{table}[t]
\caption{\label{tab:assign}Line assignments of the $pn$ channels at the central cut,
without short-range correlation: coherent rates summed over $\Lam$, $\Sig^0$ and
$\Sig^-$ with assignment $A$ alone and assignment $B$ alone, and their interference $I$,
as fractions of the complete rate $\sigma_{A+B}$; and $\sigma_\Lam/\sigma_{\Sig^0}$ for
each. Five replicas of $2.5\times10^4$ samples paired by seed, with the standard error of
the mean on the last digits in parentheses where it is at least $5\times10^{-4}$.}
\begin{ruledtabular}
\begin{tabular}{lcccccc}
 & & & & \multicolumn{3}{c}{$\sigma_\Lam/\sigma_{\Sig^0}$} \\
\cline{5-7}
$E_{\nubar}$ [GeV] & $\sigma_A/\sigma_{A+B}$ & $\sigma_B/\sigma_{A+B}$ & $I/\sigma_{A+B}$ & $A$ & $B$ & $A+B$ \\
\hline
0.6 & 0.243(1)  & 0.906(1)  & $-0.149(1)$  & 4.36(2)  & 4.67(2)  & 3.95(1) \\
1.0 & 0.254(1)  & 0.895     & $-0.149(1)$  & 4.60(5)  & 3.50(1)  & 3.30(1) \\
2.0 & 0.269(1)  & 0.889(4)  & $-0.158(3)$  & 5.20(14) & 3.32(6)  & 3.27(4) \\
3.0 & 0.272(2)  & 0.888(1)  & $-0.160(2)$  & 5.20(9)  & 3.29(6)  & 3.26(6) \\
5.0 & 0.265(10) & 0.905(6)  & $-0.169(10)$ & 5.31(37) & 3.31(18) & 3.28(13) \\
\end{tabular}
\end{ruledtabular}
\end{table}

\subsection{Hyperon composition}
\label{sec:composition}

The values below are for the truncated current of Sec.~\ref{sec:constraints}, with no
short-range correlation and at the central matching prescription. The seagull carries
the axial coupling directly and would make the current strongly $\Lam$-rich were it the
whole current ($|{-}(D+3F)/\sqrt6|=0.900$ against $(D-F)/\sqrt2=0.238$), but it is $9\%$
of the $\Lam$-channel incoherent sum at $1\GeV$; the composition is set by the poles,
which contain the one-body weak vertex, and by the decuplet.

\begin{figure}[t]
\centering
\includegraphics[width=3.4in]{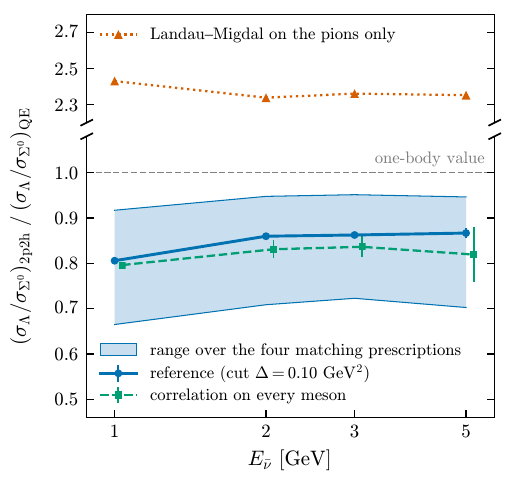}
\caption{\label{fig:composition}$\sigma_\Lam/\sigma_{\Sig^0}$ of the two-body current
divided by its one-body value in the same machinery, against antineutrino energy.
Circles: central cut, $\Delta=0.10\GeV^2$, no short-range correlation. Band: range over
the four matching prescriptions of Table~\ref{tab:hierpresc}. Circles and band are
means of $48$ replicas of $2\times10^4$ samples. Squares: the correlation of
Eq.~\eqref{eq:src} on every exchanged meson. Triangles (upper panel): the
Landau--Migdal contact of Eq.~\eqref{eq:vl} on the pions only, $g'=0.7$. Squares and
triangles are means of three replicas of $2.5\times10^4$ samples. Error bars on the
circles and squares are the standard error of the replica mean; that of the triangles,
$1$--$8\%$, is not drawn. The dashed line is the one-body value.}
\end{figure}

The two-body current is less $\Lam$-rich than the one-body one:
$\sigma_\Lam/\sigma_{\Sig^0}=3.92(1)$, $3.32(1)$, $3.30(2)$, $3.33(3)$ and $3.41(5)$ at
$0.6$, $1$, $2$, $3$ and $5\GeV$ ($48$ replicas of $2\times10^4$ samples), against
$4.12$, $3.83$, $3.86$ and $3.93$ for the one-body current in the same machinery at $1$,
$2$, $3$ and $5\GeV$. The ratio of the two, $0.81$ at $1\GeV$ and $0.86$--$0.87$ from
$2$ to $5\GeV$, is shown against energy in Fig.~\ref{fig:composition} with the band of
the matching prescriptions and the two correlated variants; the weights are
$\Lam:\Sig^0:\Sig^-\simeq3.3:1:1.2$. These ratios belong to the isolated two-body
current; in the inclusive hyperon composition they combine with its interference with
the one-body amplitude, which is not computed (Sec.~\ref{sec:interference}).

\paragraph{Axial contact sensitivity.} The axial contact operator and its kaon-pole
partner, $A_{\rm ga}+A_{{\rm ga},K}$ of Sec.~\ref{sec:constraints}, are inserted together
with strength $g$, $g=-1$ being the coefficient of Eq.~\eqref{eq:lagrangian}; $A_{\rm ip}$
and $A_{\rm sb}$ stay omitted. With common random numbers, four seed-paired replicas of
$8\times10^3$ samples give at $g=-1$ a change of $\sigma_\Lam/\sigma_{\Sig^0}$ of
$-5.6(1)\%$ at $1\GeV$ and $-1.7(5)\%$ at $3\GeV$, and of the hyperon rate of the $pn$
channels of $-1.4(2)\%$ and $+0.6(6)\%$.
The partner alone changes the rates by at most $0.1\%$, being proportional to $q^\mu$:
it restores the chiral-limit identity without changing the shifts, which are those of
the contact operator. The shift of the composition is smaller than the spread across the
matching prescriptions.

The flavour structure does not fix the sign of the departure from the quasi-elastic
composition, which depends on propagators, relative phases and the short-range
treatment: in the benchmark the two-body current is less $\Lam$-rich than the one-body
one under every prescription, and pion-only screening reverses this
(Sec.~\ref{sec:screening}). A generator should carry an independent hyperon composition:
the earlier implementation split a struck proton evenly between $\Lam$ and $\Sig^0$,
against $3.32$ here and $4.12$ for the one-body current at $1\GeV$.

\paragraph{Composition under each prescription.} Every entry of
Table~\ref{tab:comppresc} lies below the one-body value, at $0.67$--$0.95$ of it, and
the ordering follows the shares of the seagull and the decuplet, which the prescriptions
leave unregulated: at $1\GeV$ from $0.92$ of the one-body value at $\Delta=0.216\GeV^2$,
where the two carry $14\%$ and $42\%$ of the $\Lam$-channel incoherent sum, to $0.67$ at
the principal value ($4\%$ and $12\%$). Tested pairwise across shared seeds, every pair
of prescriptions is resolved at $1$ and $3\GeV$.

\begin{table}[t]
\caption{Hyperon composition under each current-to-transport prescription and with
the decuplet removed, at two energies. ``Ratio to QE'' is
$(\sigma_\Lam/\sigma_{\Sig^0})_{2p2h}$ divided by its one-body value in the same
machinery ($4.12$ at $1\GeV$, $3.86$ at $3\GeV$). Means of $48$ replicas of
$2\times10^4$ samples, with the standard error of the mean on the last digits in
parentheses. The $\Sig^+$ column is the
ordered-line share, the incoherent sum of the two orderings of the $pp$ pair without
their exchange interference, relative to the three $pn$ channels and $\Sig^+$, the
like-pair channels not included; the antisymmetrised seven-channel values are in
Sec.~\ref{sec:channels}.}
\label{tab:comppresc}
\begin{ruledtabular}
\begin{tabular}{lcccc}
prescription & $E_{\nubar}$ [GeV] & $\sigma_\Lam/\sigma_{\Sig^0}$ & ratio to QE
 & $\Sig^+$ share (ordered-line) \\
\hline
cut, $\Delta=0.046$ & 1 & 2.840(7)   & 0.689(2)  & $1.41\%$ \\
                    & 3 & 2.863(25)  & 0.741(6)  & $1.48\%$ \\
cut, $\Delta=0.10$  & 1 & 3.321(6)   & 0.806(2)  & $1.31\%$ \\
                    & 3 & 3.332(29)  & 0.862(7)  & $1.40\%$ \\
cut, $\Delta=0.216$ & 1 & 3.779(10)  & 0.917(2)  & $1.57\%$ \\
                    & 3 & 3.675(36)  & 0.951(9)  & $1.50\%$ \\
PV, $\epsilon=0.03$ & 1 & 2.741(6)   & 0.665(2)  & $1.43\%$ \\
                    & 3 & 2.793(26)  & 0.723(7)  & $1.50\%$ \\
no decuplet         & 1 & 3.061(7)   & 0.743(2)  & $1.35\%$ \\
                    & 3 & 2.756(23)  & 0.713(6)  & $1.73\%$ \\
\end{tabular}
\end{ruledtabular}
\end{table}

\subsection{Vector--axial decomposition}
\label{sec:va}

Splitting the squared amplitude into its $VV$, $AA$ and interference parts, a split exact
by construction and verified to close to $1.00000$ channel by channel, gives
Table~\ref{tab:va}. The interference is destructive in both channels, as for the
one-body current, and larger in the $\Lam$ channel, the more axial of the two: at
$1\GeV$ its $VV$, $AA$ and $VA$ parts are each of the order of the total, so the rate is
a cancellation. A relative change $\delta$ in one part, common to both channels, moves
the ratio by at most $0.27\,\delta$ at $1\GeV$ and $0.17\,\delta$ at $3\GeV$.

\begin{table}[t]
\caption{\label{tab:va}$VV$, $AA$ and $VA$ parts of the squared amplitude as fractions
of each channel's coherent total, at the central prescription. The $1\GeV$ entries are
means of eight seeds at $6\times10^3$ samples, the $3\GeV$ entries one seed at
$8\times10^3$.}
\begin{ruledtabular}
\begin{tabular}{lcccccc}
 & \multicolumn{3}{c}{$1\GeV$} & \multicolumn{3}{c}{$3\GeV$} \\
channel & $VV$ & $AA$ & $VA$ & $VV$ & $AA$ & $VA$ \\
\hline
$\Lam$   & $97\%$  & $117\%$ & $-113\%$ & $52\%$ & $69\%$ & $-21\%$ \\
$\Sig^0$ & $103\%$ & $90\%$  & $-93\%$  & $65\%$ & $52\%$ & $-17\%$ \\
\end{tabular}
\end{ruledtabular}
\end{table}

The vector current enters the ratio through $VV$ and $VA$, and the dressed vector current
is not conserved (Sec.~\ref{sec:ward}). Replacing the assembled vector current by its
transverse projection, $V^\mu\to V^\mu-q^\mu(q\cdot V)/q^2$, removes the non-conserved
longitudinal component but is not a physical restoration; it moves the ratio by $0.6\%$
at $1\GeV$ and less at $3\GeV$.
The longitudinal part of a hadronic current reaches the amplitude only through
$q_\mu\ell^\mu\propto m_\ell$, so a non-conserved longitudinal component is
$m_\ell$-suppressed in a muon-channel rate. The projection does not reach the transverse
current, which a dressing that weights Ward-related terms differently also distorts; the
Ward-compatible dressing of Sec.~\ref{sec:ward}, which changes both, moves the ratio by
at most $4.3\%$ (Table~\ref{tab:dressscheme}).

\subsection{Decuplet weight}
\label{sec:decuplet}

The weight of the decuplet class is measured in five ways. \emph{(i) Incoherent family share:}
$26$--$48\%$ of the $\Lam$ channel at $1$--$5\GeV$ (Table~\ref{tab:classes}) and
$24$--$42\%$ summed over the three hyperon channels. \emph{(ii) Coherent removal,}
$1-\sigma_{\rm no\,dec}/\sigma_{\rm all}$ on the same random numbers: $35$--$55\%$ of the
channel-summed $\DS=1$ strength at $1$--$5\GeV$. \emph{(iii) Widths:} broadening the
$\Sst$ width to $61.5\MeV$, in proportion to an in-medium $\Delta$ width of $200\MeV$
that the crossed $\Delta$ also takes, with the couplings held at the free widths, lowers
the coherent $\DS=1$ rate by up to $23\%$ over $0.6$--$5\GeV$, and a $\Sst$ width of
$100\MeV$ alone by up to $35\%$ (Table~\ref{tab:dsfactors}). \emph{(iv) Composition:}
deleting the class, with the replicas paired by seed, changes
$\sigma_\Lam/\sigma_{\Sig^0}$ by $+2\%$ at $0.6\GeV$ and by $-8\%$ to $-17\%$ at
$1$--$5\GeV$: above $1\GeV$ the decuplet makes the current more $\Lam$-rich, and without
it the ratio to the quasi-elastic value is $0.74$ at $1\GeV$ and $0.71$ at $3\GeV$
(Table~\ref{tab:comppresc}). \emph{(v) Phase:} reversing the whole class against the
octet families, as a diagnostic of the sign that SU(3) and the Goldberger--Treiman
relations fix (Sec.~\ref{sec:classes}), lowers $\sigma_\Lam/\sigma_{\Sig^0}$ by
$15$--$20\%$ at every energy and the channel-summed rate by $2$--$11\%$.

\subsection{Off-shell sensitivity}
\label{sec:offshell}

For short-range treatments that act on every exchanged meson alike, the separation
between the off-shell baryon-pole current and an on-shell cascade step is the dominant
uncertainty of the direct evaluation. The quantity it moves is a scheme-dependent pair
strength: over the window $\Delta=0.046$--$0.216\GeV^2$ the coherent $\DS=1$ pair
strength, summed over channels, changes by a factor $3.4$ at $1\GeV$ and by $2.3$--$6.7$
over $0.6$--$5\GeV$, most at $0.6\GeV$ and in the channels where the regulated poles
carry most of the strength. Table~\ref{tab:dsfactors} gives the factors for every
prescription and for the other variants run on the $\DS=1$ rate.

\paragraph{Regulated principal value.} The principal-value entries of
Tables~\ref{tab:hierpresc} and~\ref{tab:comppresc} use $\epsilon=0.03\GeV^2$, which
retains $2.0$--$2.4$ times the coherent $\DS=1$ pair strength of the central cut over
$1$--$5\GeV$, against $0.59$--$2.03$ across the cut window; the composition rises with
$\epsilon$, from $2.58$ at $\epsilon=0.01\GeV^2$ to $3.59$ at $0.10\GeV^2$ at $1\GeV$.
The cut and the principal value are distinct matching schemes, and a normalisation
built on this current carries the choice between them in addition to the scale
dependence within each.

\begin{table}[t]
\caption{\label{tab:dsfactors}Coherent $\DS=1$ rate, summed over the $\Lam$, $\Sig^0$
and $\Sig^-$ channels, under each variant, relative to the reference evaluation (central
cut $\Delta=0.10\GeV^2$, no short-range correlation, $|\mathbf q|<1.2\GeV$, the inputs
of Table~\ref{tab:inputs}). Each entry is the mean over replicas of the ratio paired by
seed, with the standard error of the mean on the last digit in parentheses where it is
at least $5\times10^{-4}$: five replicas of $2\times10^4$ samples for the cuts and for
$\epsilon=0.03\GeV^2$, three for the $|\mathbf q|$ variants, one for $D$ and $F$, and two
otherwise. The uniform correlation at $q_c=0.78\GeV$ and the
pion-only contact are in Table~\ref{tab:screening}.}
\begin{ruledtabular}
\begin{tabular}{l*{5}{d{7}}}
variant & \multicolumn{1}{c}{$0.6\GeV$} & \multicolumn{1}{c}{$1\GeV$} & \multicolumn{1}{c}{$2\GeV$} & \multicolumn{1}{c}{$3\GeV$} & \multicolumn{1}{c}{$5\GeV$} \\
\hline
cut, $\Delta=0.046\GeV^2$ & 2.63(1) & 2.03(1) & 1.83(4) & 1.74(3) & 1.74(5) \\
cut, $\Delta=0.216\GeV^2$ & 0.394(3) & 0.595(3) & 0.663(1) & 0.707(14) & 0.755(19) \\
PV, $\epsilon=0.02\GeV^2$ & 5.14(3) & 3.60(2) & 3.08(12) & 2.89(15) & 2.73(26) \\
PV, $\epsilon=0.03\GeV^2$ & 3.26(2) & 2.43(2) & 2.07(4) & 2.00(6) & 2.06(15) \\
PV, $\epsilon=0.06\GeV^2$ & 1.44(1) & 1.26 & 1.22(2) & 1.19(2) & 1.16(4) \\
\hline
uniform correlation, $q_c=0.60\GeV$ & 0.241(1) & 0.222(1) & 0.209(1) & 0.207(1) & 0.202(4) \\
uniform correlation, $q_c=1.00\GeV$ & 0.459 & 0.445(1) & 0.433(1) & 0.432(2) & 0.427(4) \\
\hline
decuplet deleted & 0.862 & 0.651(2) & 0.526(6) & 0.528(24) & 0.446(13) \\
$\Gamma_{\Sig^*}=61.5\MeV$, $\Gamma_\Delta=200\MeV$ & 0.979(1) & 0.875(2) & 0.814(2) & 0.820(17) & 0.774(10) \\
$\Gamma_{\Sig^*}=100\MeV$ & 0.961(1) & 0.803(3) & 0.709(4) & 0.717(24) & 0.652(14) \\
\hline
$|\mathbf q|<1.1\GeV$ & 1.000 & 0.983 & 0.946(1) & 0.938(1) & 0.942(4) \\
$|\mathbf q|<1.3\GeV$ & 1.000 & 1.010 & 1.035 & 1.051(1) & 1.064(11) \\
no $|\mathbf q|$ restriction & 1.000 & 1.013 & 1.235(3) & 1.459(13) & 1.837(61) \\
\hline
$M_A=0.9\GeV$ & 0.970 & 0.975 & 0.977 & 0.977 & 0.975(4) \\
$M_A=1.2\GeV$ & 1.057 & 1.055 & 1.047 & 1.045 & 1.048(7) \\
$\Lambda_\pi=\Lambda_K=1.0\GeV$ & 0.850(1) & 0.823(1) & 0.805(2) & 0.804(3) & 0.795(6) \\
$\Lambda_\pi=\Lambda_K=1.5\GeV$ & 1.066 & 1.083(1) & 1.095(1) & 1.095(2) & 1.101(3) \\
$D=0.812$ & 1.005 & 1.003 & 1.003 & 1.003 & 1.002 \\
$D=0.796$ & 0.996 & 0.997 & 0.998 & 0.997 & 0.998 \\
$F=0.475$ & 1.003 & 1.002 & 1.001 & 1.001 & 1.001 \\
$F=0.459$ & 0.997 & 0.998 & 0.999 & 0.999 & 0.999 \\
\end{tabular}
\end{ruledtabular}
\end{table}

\subsection{Correlation variants}
\label{sec:screening}

The two correlated treatments of Sec.~\ref{sec:srcdef} are run at the central cut with
everything else unchanged: the Oset--Toki--Weise correlation of Eq.~\eqref{eq:src} with
$q_c=0.78\GeV$ on every exchanged meson (``uniform''), and the Landau--Migdal contact of
Eq.~\eqref{eq:vl} with $g'=0.7$ on every exchanged pion, the $K$ and $\eta$ legs bare
(``$\pi$ only''). Table~\ref{tab:screening} gives their effect on the composition and on
the coherent $\DS=1$ rate.

\begin{table}[t]
\caption{\label{tab:screening}$\sigma_\Lam/\sigma_{\Sig^0}$, and the coherent $\DS=1$
rate summed over the $\Lam$, $\Sig^0$ and $\Sig^-$ channels relative to the reference,
$\sigma/\sigma_{\rm ref}$, with no correlation (reference), the uniform correlation of
Eq.~\eqref{eq:src} and the pion-only contact of Eq.~\eqref{eq:vl}, at the central cut.
QE is the one-body value of $\sigma_\Lam/\sigma_{\Sig^0}$. For
$\sigma_\Lam/\sigma_{\Sig^0}$ the reference column is the mean of $48$ replicas of
$2\times10^4$ samples and the uniform and $\pi$-only columns are means of three
replicas of $2.5\times10^4$, with the standard error of the mean over the replicas on
the last digits in parentheses; the seed-paired changes are in the text.
$\sigma/\sigma_{\rm ref}$ is the mean of the seed-paired ratio over four
replicas of $2\times10^4$ samples.}
\begin{ruledtabular}
\begin{tabular}{lcccccc}
 & \multicolumn{4}{c}{$\sigma_\Lam/\sigma_{\Sig^0}$} & \multicolumn{2}{c}{$\sigma/\sigma_{\rm ref}$} \\
\cline{2-5}\cline{6-7}
$E_{\nubar}$ [GeV] & QE & reference & uniform & $\pi$ only & uniform & $\pi$ only \\
\hline
0.6 & ---  & 3.92(1)  & 3.93(5)  & 13.3(1) & 0.348(1) & 0.078(1) \\
1.0 & 4.12 & 3.32(1)  & 3.28(1)  & 10.0(1) & 0.330(1) & 0.091(1) \\
2.0 & 3.83 & 3.30(2)  & 3.19(8)  & 9.0(2)  & 0.318(1) & 0.091(2) \\
3.0 & 3.86 & 3.33(3)  & 3.23(9)  & 9.1(1)  & 0.314(1) & 0.093(4) \\
5.0 & 3.93 & 3.41(5)  & 3.22(24) & 9.3(7)  & 0.309(2) & 0.096(1) \\
\end{tabular}
\end{ruledtabular}
\end{table}

\paragraph{Uniform correlation.} The correlation lowers the coherent $\DS=1$ rate by a
factor of about three at every energy, nearly equally in the three channels, and
changes the composition by under $3\%$ (seed-paired, $-0.4$ to $-2.1\%$ over
$0.6$--$5\GeV$). It suppresses kaon exchange more than the other
families, by $4.4$--$6.1$ against $2.9$--$3.3$, and leaves the pole shares nearly
unchanged. At $5\GeV$ the squares of Fig.~\ref{fig:composition} lie below the circles
because the three reference replicas they are paired with give a lower
$\sigma_\Lam/\sigma_{\Sig^0}$ than the mean of all $48$.

\paragraph{Pion-only Landau--Migdal contact.} At the exchanged momenta of a correlated
pair the contact cancels most of the pion pole, while the kaon-exchange family keeps its
bare propagator. That family becomes $53\%$ of the $\Lam$-channel incoherent sum at
$1\GeV$ and $41\%$ at $3\GeV$, the decuplet, whose line-2 $\Sst$ also exchanges a kaon,
$27\%$ and $45\%$, and the two pole orderings fall to $11\%$ and $8\%$: at $1\GeV$ the
leading family is one whose range is $1/m_K$. Kaon absorption on the hyperon line
carries $g_{KN\Lam}\propto D+3F$ against $g_{KN\Sig}\propto D-F$, so a current dominated
by it is $\Lam$-rich: $\sigma_\Lam/\sigma_{\Sig^0}$ rises by a factor $2.75$--$3.36$
over $0.6$--$5\GeV$, to $2.3$--$2.4$ times the quasi-elastic value over $1$--$5\GeV$. The
response to $g'$ is not monotonic, the contact first cancelling and then exceeding the
pion pole: at $1\GeV$ the ratio peaks at $g'=0.8$ and falls at $0.9$. The coherent
$\DS=1$ rate falls by a factor $10$--$13$, by $8$--$9$ in the $\Lam$ channel, which keeps
its bare kaon exchange, and by $22$--$34$ in the $\Sig$ channels. Since the correlation
hole belongs to the $NN$ pair, this variant measures the effect of screening the
exchanged mesons asymmetrically, and the uniform one the effect of adding a correlation
to the uncorrelated reference.

\subsection{Distributions}
\label{sec:distributions}

\begin{figure*}[t]
\centering
\includegraphics[width=0.92\textwidth]{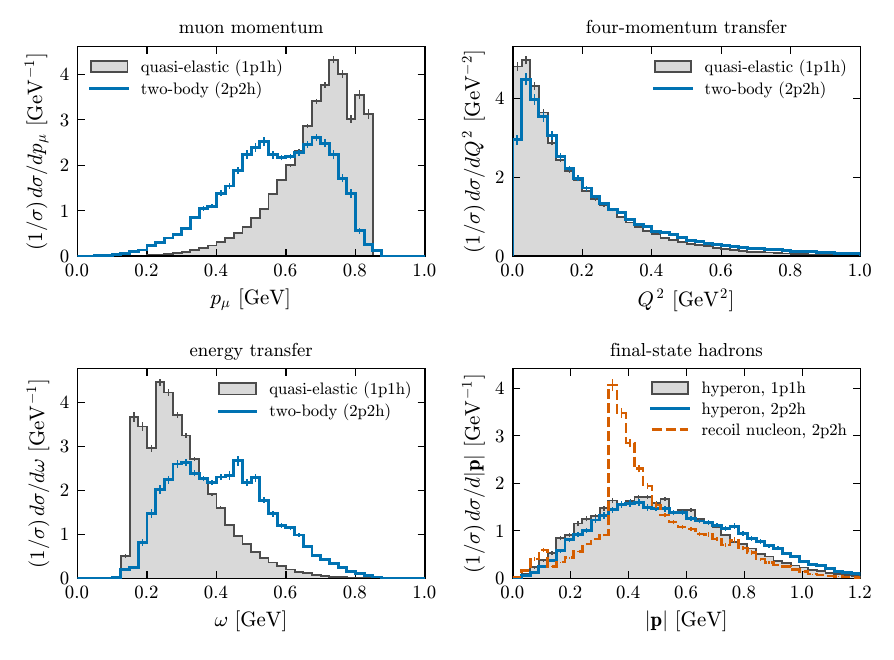}
\caption{\label{fig:kinematics}Quasi-elastic (1p1h) and two-body (2p2h) $\DS=1$
production at $E_{\nubar}=1\GeV$, summed over $\Lam$, $\Sig^0$ and $\Sig^-$: muon
momentum, four-momentum transfer, energy transfer, and final-state hadron momenta.
Both currents share the machinery, the struck-nucleon momentum and
$|\mathbf q|<1.2\GeV$; the two-body current is at the central cut
$\Delta=0.10\GeV^2$ without short-range correlation. Each distribution is normalised
to unit area; the quasi-elastic one is the filled histogram. Error bars are the Monte
Carlo errors.}
\end{figure*}

Figure~\ref{fig:kinematics} compares the two currents at $E_{\nubar}=1\GeV$, with every
struck nucleon at fixed $|\mathbf p|=0.25\GeV$ (the pair back to back at zero total
momentum) and $|\mathbf q|<1.2\GeV$ applied to both; the shapes come from the factorised
pair of Sec.~\ref{sec:factorised}. Sharing the energy with a second nucleon lowers the
muon momentum and raises $\omega$ and $Q^2$, as in the non-strange 2p2h response
(Table~\ref{tab:means}); the two-body energy-transfer distribution overlaps the
quasi-elastic one but reaches $0.9\GeV$. The recoil nucleon is softer than the hyperon:
except in kaon exchange, the momentum transfer enters on the hyperon line and the recoil
receives only the exchanged-meson momentum.

\begin{table}[b]
\caption{\label{tab:means}Mean values of the distributions of Fig.~\ref{fig:kinematics}.}
\begin{ruledtabular}
\begin{tabular}{lrr}
 & 1p1h & 2p2h \\
\hline
$\langle p_\mu\rangle$ [GeV]      & 0.69 & 0.57 \\
$\langle Q^2\rangle$ [GeV$^2$]    & 0.17 & 0.21 \\
$\langle \omega\rangle$ [GeV]     & 0.30 & 0.42 \\
$\langle |\mathbf{p}_Y|\rangle$ [GeV] & 0.51 & 0.55 \\
$\langle |\mathbf{p}_{N'}|\rangle$ [GeV] & --- & 0.48 \\
\end{tabular}
\end{ruledtabular}
\end{table}

The sharp maximum of the recoil spectrum at $0.34\GeV$ is an edge of the matching
prescription: with the pair at rest the denominator of the absorption-first nucleon
pole, Eq.~\eqref{eq:nucpole}, is $4E_N(E_N-E_{N'})$ and vanishes when
$|\mathbf p_{N'}|$ equals the pair-nucleon momentum, so the central cut sets that term
to zero for $0.11<|\mathbf p_{N'}|<0.34\GeV$. Without Pauli blocking of the recoil, the
soft end of its spectrum is overstated. The recoil distribution depends on the pair
model and the matching prescription, and a detector study would need the unfactorised
response.

\subsection{Charge channels}
\label{sec:channels}

The evaluation carries all seven channels of Table~\ref{tab:channels}: the three $pn$
channels in both line assignments, and the $pp$ and $nn$ channels with the identical
initial nucleons antisymmetrised,
$\mathcal M_{\rm as}=\mathcal M(p_1s_1,p_2s_2)-\mathcal M(p_2s_2,p_1s_1)$
(Sec.~\ref{sec:2p2hresponse}). The exchange term is a different kinematic
configuration, the meson now flowing from the other nucleon, re-assembled on the swapped
kinematics at every phase-space point and combined with the direct term before squaring.
Table~\ref{tab:antisym} gives the ratio
$\langle|\mathcal M_{\rm as}|^2\rangle/(2\langle|\mathcal M(1,2)|^2\rangle)$, unity
when the exchange interference vanishes, and the share of the pair spin singlet in
$\sum|\mathcal M_{\rm as}|^2$.

\begin{table}[t]
\caption{\label{tab:antisym}Antisymmetrisation of the $pp\to\Sig^+n$ amplitude. Third
column: $\langle|\mathcal M_{\rm as}|^2\rangle/(2\langle|\mathcal M(1,2)|^2\rangle)$ with
$\mathcal M_{\rm as}=\mathcal M(1,2)-\mathcal M(2,1)$, averaged over the
normalised symmetric pair distribution; it is unity when exchange interference
vanishes. Fourth column: the spin-singlet share of $\sum|\mathcal M_{\rm as}|^2$ summed
over final and lepton spins; the remainder is the triplet. Last column: the $\Sig^+$
share of the isolated two-body hyperon rate summed over all seven channels, identical pairs
antisymmetrised, in the reference pair census. Central rows: $16$ replicas of
$2\times10^4$ samples, standard error of the mean on the last digits in parentheses where
it is at least half a unit; other rows: three replicas.}
\begin{ruledtabular}
\begin{tabular}{lcccc}
prescription & $E_{\nubar}$ [GeV] & $\langle|\mathcal M_{\rm as}|^2\rangle/(2\langle|\mathcal M(1,2)|^2\rangle)$ & $S=0$ share & $\Sig^+$ share [\%] \\
\hline
cut, $\Delta=0.10$ & $0.6$ & $0.970(2)$ & $29\%$ & $1.135(2)$ \\
                       & $1.0$ & $0.917(2)$ & $29\%$ & $1.153(2)$ \\
                       & $2.0$ & $0.887(4)$ & $29\%$ & $1.187(7)$ \\
                       & $3.0$ & $0.888(7)$ & $29\%$ & $1.196(14)$ \\
                       & $5.0$ & $0.880(16)$ & $29\%$ & $1.178(24)$ \\
\hline
cut, $\Delta=0.046$    & $1.0$ & $0.950(2)$ & $26\%$ & $1.292(11)$ \\
                       & $3.0$ & $0.930(22)$ & $27\%$ & $1.317(55)$ \\
cut, $\Delta=0.216$    & $1.0$ & $0.834(2)$ & $32\%$ & $1.289(7)$ \\
                       & $3.0$ & $0.833(11)$ & $31\%$ & $1.199(17)$ \\
PV, $\epsilon=0.03$    & $1.0$ & $0.948(5)$ & $26\%$ & $1.290(16)$ \\
                       & $3.0$ & $0.924(39)$ & $26\%$ & $1.334(44)$ \\
\end{tabular}
\end{ruledtabular}
\end{table}

\begin{table}[t]
\caption{\label{tab:likepair}Channels of identical pairs at the central prescription:
antisymmetrised rate per pair relative to the $pn$ channel with the same hyperon
($pn\to\Lam n$, $pn\to\Sig^0n$, $pn\to\Sig^-p$, both line assignments coherent), and the
antisymmetrisation ratio
$R=\langle|\mathcal M_{\rm as}|^2\rangle/(2\langle|\mathcal M(1,2)|^2\rangle)$. $16$
replicas of $2\times10^4$ samples; standard errors on the last digits in parentheses.}
\begin{ruledtabular}
\begin{tabular}{l*{3}{d{9}}*{3}{d{7}}}
 & \multicolumn{3}{c}{rate per pair / same-hyperon $pn$} & \multicolumn{3}{c}{$R$} \\
\cline{2-4}\cline{5-7}
$E_{\nubar}$ [GeV] & \multicolumn{1}{c}{$pp\to\Lam p$} & \multicolumn{1}{c}{$pp\to\Sig^0p$} & \multicolumn{1}{c}{$nn\to\Sig^-n$} & \multicolumn{1}{c}{$\Lam p$} & \multicolumn{1}{c}{$\Sig^0p$} & \multicolumn{1}{c}{$\Sig^-n$} \\
\hline
0.6 & 0.5928(10) & 0.4422(13)  & 0.7834(14)  & 1.081(2)  & 0.964(2)  & 1.058(2) \\
1.0 & 0.5221(15) & 0.3632(13)  & 0.9314(19)  & 0.963(2)  & 0.945(3)  & 0.978(2) \\
2.0 & 0.4909(34) & 0.3298(40)  & 1.0446(57)  & 0.905(4)  & 0.909(5)  & 0.932(5) \\
3.0 & 0.4830(38) & 0.3259(43)  & 1.0605(74)  & 0.900(10) & 0.910(10) & 0.931(10) \\
5.0 & 0.4738(71) & 0.3357(133) & 1.0701(229) & 0.903(10) & 0.935(33) & 0.924(16) \\
\end{tabular}
\end{ruledtabular}
\end{table}

The $\Sig^+$ channel survives at every energy and prescription. The exchange
interference is destructive, the antisymmetrised rate being $0.97$ of the ordered-line
one at $0.6\GeV$, $0.88$--$0.92$ of it at $1$--$5\GeV$ and $0.83$--$0.95$ across the
prescriptions. The singlet carries $29\%$, above its statistical weight of $1/4$; the
rest is triplet, which requires a spatially antisymmetric amplitude. In the other
like-pair channels (Table~\ref{tab:likepair}) the exchange interference is constructive
only at $0.6\GeV$, for $pp\to\Lam p$ and $nn\to\Sig^-n$, and destructive otherwise.
These shares characterise the factorised free-pair amplitude, not nuclear pair-state
probabilities, since no partial-wave-resolved two-body density enters.

Weighted by the reference pair census, the $pp\to\Lam p$, $\Sig^0p$ and $nn\to\Sig^-n$
channels add $3.8\%$ to the rate of the $pn$ channels and $\Sig^+$ and change
$\sigma_\Lam/\sigma_{\Sig^0}$ by at most $0.9\%$. The antisymmetrised $\Sig^+$ share of
the isolated two-body hyperon rate, all seven channels in the denominator, is
$1.13$--$1.20\%$ over $0.6$--$5\GeV$ and $1.20$--$1.33\%$ across the prescriptions at $1$
and $3\GeV$. It depends less on the short-range treatment than the composition does:
the uniform correlation leaves it unchanged and the pion-only contact lowers it by
$27$--$42\%$, while $\sigma_\Lam/\sigma_{\Sig^0}$ roughly triples.

\paragraph{Pair census.} The share carries a nuclear input alongside the current, the
pair weights that set the $pp$ requirement at the rate level. Counting pairs in
$^{40}$Ar ($Z=18$, $N=22$) gives $153$ $pp$, $231$ $nn$ and $396$ $pn$ combinations. The
reference pair-weight model takes the $pn$ dominance measured in short-range-correlation
studies~\cite{subedi2008,hen2017} as a ratio of pair numbers, $N_{pn}/N_{pp}\simeq18$,
with the like pairs in their combinatorial ratio, $N_{nn}/N_{pp}=231/153$: $87.8\%$ $pn$,
$4.9\%$ $pp$, $7.4\%$ $nn$. The $\Sig^+$ rate relative to the $pn$ channels scales as
$N_{pp}/N_{pn}$: over $N_{pn}/N_{pp}=12$--$24$ the share runs from $1.69$ to $0.87\%$ at
$1\GeV$.
The $pn$ dominance is measured for high-relative-momentum pairs in particular kinematics
and is $A$-dependent, so the census is a model input standing in for a momentum- and
spin-dependent two-body density.

\paragraph{The kaon-less $\Sigma^+$ tag.} What this rate implies for a kaon-less
$\Sig^+$ tag, a sample in-medium $YN$ analyses attribute entirely to final-state
interactions, is a generator-level question~\cite{Nowak:2026pheno}: it divides this rate
by a predicted final-state-interaction yield and inherits that model's dependence.
Final-state interactions are estimated to produce a kaon-less $\Sig^+$ in about $1\%$
of antineutrino quasi-elastic hyperon events~\cite{companionPRD}. In an additive
treatment that neglects the one-body--two-body interference, as event generators do, the
direct channel supplies at most about $1.2\%$ of hyperon events in the reference census
at any two-body normalisation. With the two-body rate at $0.14$--$0.22$ times the
quasi-elastic hyperon rate~\cite{Nowak:2026pheno}, two-body events are about $11\%$ of
the hyperon events of a generated antineutrino-mode sample in the Booster Neutrino
Beam, the flux and the generator's own quasi-elastic rate folded in, and
primary $\Sig^+$ that leave the nucleus are $11$--$12\%$ of its kaon-less $\Sig^+$ tag.

\section{Limitations}
\label{sec:notdetermined}
\label{sec:outlook}

\paragraph{Nuclear response and normalisation.} The factorised treatment yields a cross
section per correlated pair, of dimension $\GeV^{-4}$, because it replaces the density
factor of the $\int d^3p_2\,n(p_2)$ integral of Eq.~\eqref{eq:2p2h} by a normalised pair
distribution at zero total momentum; no absolute $\DS=1$ 2p2h cross section is
therefore quoted. Evaluating Eq.~\eqref{eq:2p2h} with the local Fermi gas and the LDA
integral, for which the class operators derived here are the input, would supply that
factor, and with it the correlation between the recoil direction and the momentum
transfer that would let an experiment tag this channel (its non-strange analogue is
quantified in Ref.~\cite{kasturi2026}). The in-medium resonance
self-energy~\cite{osetsalcedo1987,kaskulov2006}, whose strange half is unmeasured,
bears on the absolute normalisation rather than on the ratios. Three-nucleon $\DS=1$
terms, the analogue of $\Lam NN\to NNN$~\cite{itonaga2023} and of the three-body
absorption of non-strange calculations~\cite{nieves2011,sobczyk2020}, are not
included.

\paragraph{Interference with the one-body current.} The one- and two-body currents share
final states, and nothing requires their interference to weight $\Lam$ and $\Sig^0$
equally. It is neither computed nor bounded in a factorised treatment
(Sec.~\ref{sec:interference}), so the composition quoted is that of the two-body current
in isolation.

\paragraph{Matching and short-range treatment.} Screening the pions alone hands the
$\Lam$-channel current to kaon exchange and the decuplet (Sec.~\ref{sec:screening}), so a
nuclear calculation of the strange response needs a short-range treatment that includes
the kaon exchange. The matching dependence can be removed only by treating the current
together with the hyperon cascade that transports the final state.

\paragraph{Open pieces of the current.} The evaluated current lacks the transverse
off-shell completion and the axial terms $A_{\rm ga}$, $A_{{\rm ga},K}$, $A_{\rm ip}$ and
$A_{\rm sb}$ (Sec.~\ref{sec:constraints}). Its reference form factors break the vector
identity at a median residual of $0.27$, and the Ward-compatible dressing of
Sec.~\ref{sec:ward} is one of a family of dressings that preserve it before
matching~\cite{grossriska1987,haberzettl1998}, differing by transverse structures and
by common rescalings of Ward-related sets. The decay constants are taken in the SU(3)
limit, $f_K=f_\pi$, and the decuplet carries the $\Delta$ transition form factors
transferred to the $\Sst$ by SU(3)~\cite{alam2012}, with $C_4^A$ omitted, $C_5^A$ a
pure dipole~\cite{hnv2007} and a line-2 $\Sst N\bar K$ coupling that rests on SU(3)
alone.

\section{Summary}
\label{sec:conclusions}

We have constructed the strangeness-changing two-body current for antineutrino--nucleus
scattering, $\nubar\,[NN]\to\ell^+ Y N'$, from vertices of the lowest-order SU(3) chiral
Lagrangian, summed coherently over both line assignments of a proton--neutron pair with
the relative sign of Fermi statistics. Before dressing and matching, its vector part is
the Noether current of the flavour rotation and is conserved pointwise in the SU(3)
limit, and the divergence of its axial part, $A^\mu_{\rm comp}$, equals
$(m_s+m_u)P_{su}$ chain by chain at machine precision, by construction rather than as a
test of the full axial Ward--Takahashi identity; the axial contact operator, its
kaon-pole partner and the remaining inverse-propagator and mass-splitting terms are
omitted.

The charge-exchange structure opens a direct, kaon-less $\Sig^+$ from a $pp$ pair, which
no one-body weak vertex reaches; antisymmetrised and weighted by the reference argon pair
census, it is $1.13$--$1.20\%$ of the isolated two-body hyperon rate over $0.6$--$5\GeV$,
all seven charge channels included. The $KNY$ couplings carry the axial $N\!\to\!Y$
flavour combinations, so a common scale factor across hyperon channels has no support
in the flavour structure.

In the benchmark the octet baryon poles carry $77\%$ of the $\Lam$-channel incoherent
family sum at $0.6\GeV$ and $43$--$58\%$ at $1$--$5\GeV$ at the central prescription
($30$--$88\%$ at $0.6$--$5\GeV$ across prescriptions) and the decuplet $26$--$48\%$
at $1$--$5\GeV$: pole dominance holds only near threshold and for narrow
matching scales. The meson-in-flight term stays below $1\%$, its second propagator
being a kaon. The families interfere destructively, and so do the two line assignments,
by $15$--$17\%$ of the rate. At the central prescription
$\sigma_\Lam/\sigma_{\Sig^0}=3.32$ at $1\GeV$ against $4.12$ for the one-body current;
over $1$--$5\GeV$ the isolated two-body current is $0.81$--$0.87$ times as $\Lam$-rich as
the one-body one, and its effect on the inclusive composition depends also on the
uncalculated interference with the one-body amplitude. A generator should carry the
two-body hyperon composition independently of the quasi-elastic one.

The one-body--two-body interference and the nuclear response, which sets the absolute
scale, are not evaluated. For a short-range treatment that acts on every exchanged
meson alike, the largest evaluated uncertainty is the current-to-transport matching:
across the formation-length window it changes the $\DS=1$ pair strength by a factor
$2.3$--$6.7$, and over the four prescriptions $\sigma_\Lam/\sigma_{\Sig^0}$ lies between
$0.67$ and $0.95$ of its quasi-elastic value. One correlation function on every
exchanged meson lowers the rate by a factor of about three but changes
$\sigma_\Lam/\sigma_{\Sig^0}$ by less than $3\%$; the Landau--Migdal contact on the pions
alone lowers the rate by a factor $10$--$13$, leaves the kaon exchange bare to carry
$53\%$ of the $\Lam$-channel sum at $1\GeV$, and roughly triples
$\sigma_\Lam/\sigma_{\Sig^0}$. After the matching, the decuplet input is the largest:
removing the class moves $\sigma_\Lam/\sigma_{\Sig^0}$ by up to $17\%$, and its width
the rate by up to $35\%$. The axial contact operator with its kaon-pole partner lowers
$\sigma_\Lam/\sigma_{\Sig^0}$ by $6\%$ at $1\GeV$ and $2\%$ at $3\GeV$, and the
Ward-compatible dressing changes it by about $4\%$.

\begin{acknowledgments}
This work was supported by the Science and Technology Facilities Council (STFC)
Lancaster EPP Consolidated Grant 2025--2029: UKRI2846.
\end{acknowledgments}

\appendix

\section{Code availability}
\label{sec:repro}

All numbers come from one C++17 program that evaluates the current of
Sec.~\ref{sec:twobody} with flavour tables generated symbolically from the traces of
Eq.~\eqref{eq:traces}, together with the Noether reference implementation of
Appendix~\ref{app:worked}. Monte Carlo seeds are fixed in the source; independent
replicas use distinct seeds, and paired comparisons use common random numbers. The program, the trace generator and the archived
per-replica outputs behind every table and figure are available from the author on
request.

The implementation reproduces the on-shell transversality identities of the decuplet
propagator, $p_\alpha[(\slashed p+M_R)P^{\alpha\beta}]=0$ and
$\gamma_\alpha[(\slashed p+M_R)P^{\alpha\beta}]=0$, to $10^{-16}$; the couplings of
Eq.~\eqref{eq:width} round-trip to the input widths, and the vector part of
Eq.~\eqref{eq:decvertex} agrees with the vertex of Ref.~\cite{hnv2007} to $10^{-15}$.

\section{Noether insertion sum}
\label{app:worked}

The vector Ward identity is assembled for one complete physical channel,
$\nubar\,(pn)\to\ell^+\Lam n$, the largest, in both of its line assignments. The
closure reduces to four relations among the trace coefficients
(Appendix~\ref{app:ledger}), and the reference implementation closes at
$1.2\times10^{-15}$ per spin configuration.

\subsection{The insertion sum}

In the SU(3) limit the $\DS=1$ vector current is the Noether current of
$\delta X=i[h,X]$, with $h$ the $u\!\to\!s$ source matrix, applied to
Eq.~\eqref{eq:lagrangian} and the meson kinetic term, expanded to one meson field at each
meson--baryon vertex,
\begin{equation}
  \mathcal L = \langle\bar B\,i\slashed\partial B\rangle
  - \frac{\sqrt2}{f}\Big[\frac{D}{2}\langle\bar B\gamma^\mu\gamma_5\{\partial_\mu\Phi,B\}\rangle
  + \frac{F}{2}\langle\cdots[\partial_\mu\Phi,B]\rangle\Big]
  + \frac12\langle\partial\Phi\,\partial\Phi\rangle,
\end{equation}
with the canonically normalised $\Phi$, giving
three insertion terms:
\begin{equation}
  V^\mu \;=\; -\langle\bar B\gamma^\mu[h,B]\rangle
  \;+\; i\langle\partial^\mu\Phi\,[h,\Phi]\rangle
  \;-\; \frac{i\sqrt2}{f}\Big[\frac{D}{2}\langle\bar B\gamma^\mu\gamma_5\{[h,\Phi],B\}\rangle
  + \frac{F}{2}\langle\cdots\rangle\Big].
  \label{eq:noether3}
\end{equation}
On the one-meson-exchange base diagram $p(p_1)\,n(p_2)\to\Lam(p_1')\,n(p_2')$ the first
term inserts on every baryon leg --- in-1 ($Y$ pole), out-1 ($N$ pole), in-2 (the
kaon-exchange family); the out-2 insertion vanishes because no $\DS=1$ rotation connects
$n$ to $n$ --- the second converts the exchanged meson in flight ($\pi\to K$), and the
third generates a Kroll--Ruderman seagull at each strong vertex. The same construction
with the lines exchanged, the neutron on line 1 and the partner emitting a $\pi^+$, gives
the second line assignment, a separately conserved current.

\subsection{The route list, with coefficients}

Table~\ref{tab:worked} lists every route the traces generate for this channel, with its
trace coefficient $c$ of Eq.~\eqref{eq:traces}; each vertex, seagulls included, enters
the amplitude as $\sqrt2\,c/f$ (Sec.~\ref{sec:isospin}), so every route carries the
common factor $2/f^2$, which drops out of the closure. Each entry is an exact $(D,F)$
combination; the machine-generated table used by the code agrees with every entry to
$10^{-10}$.

\begin{table}[t]
\caption{Complete route list for $pn\to\Lam n$ in both line assignments ($A$: the
weak current on the proton; $B$: on the neutron): exchange, insertion family, and the
trace coefficient $c$, quoted analytically and numerically ($D=0.804$, $F=0.467$); the
vertex is $\sqrt2\,c/f$, and assignment $B$ enters the amplitude with a minus sign.}
\label{tab:worked}
\begin{ruledtabular}
\begin{tabular}{llll}
exchange & family & coefficient & value \\
\hline
$\pi^0$ & line-2 emission ($n\to n\pi^0$)      & $-(D+F)/(2\sqrt2)$   & $-0.4494$ \\
$\pi^0$ & weak-first pole, via $\Sig^0$        & $+D/\sqrt6$          & $+0.3282$ \\
$\pi^0$ & absorption-first pole, via $p$       & $+(D+F)/(2\sqrt2)$   & $+0.4494$ \\
$\pi^0$ & Kroll--Ruderman seagull              & $-(D+3F)/(4\sqrt3)$  & $-0.3183$ \\
$\pi^0$ & meson-in-flight ($\pi^0\to K^-$)     & $+(D+3F)/(4\sqrt3)$  & $+0.3183$ \\
\hline
$\eta$  & line-2 emission ($n\to n\eta$)       & $+(3F-D)/(2\sqrt6)$  & $+0.1219$ \\
$\eta$  & weak-first pole, via $\Lam$          & $-D/\sqrt6$          & $-0.3282$ \\
$\eta$  & absorption-first pole, via $p$       & $+(3F-D)/(2\sqrt6)$  & $+0.1219$ \\
$\eta$  & Kroll--Ruderman seagull              & $-(D+3F)/4$          & $-0.5513$ \\
$\eta$  & meson-in-flight ($\eta\to K^-$)      & $+(D+3F)/4$          & $+0.5513$ \\
\hline
$K$     & line-1 absorption ($K^-p\to\Lam$)    & $-(D+3F)/(2\sqrt6)$  & $-0.4501$ \\
$K$     & line-2 weak route, via $\Sig^-$      & $+(D-F)/2$           & $+0.1685$ \\
$K$     & line-2 kaon seagull                  & $+(D-F)/2$           & $+0.1685$ \\
\hline
\multicolumn{4}{l}{assignment $B$ (the weak current on the neutron)} \\
$\pi^+$ & line-2 emission ($p\to n\pi^+$)      & $+(D+F)/2$           & $+0.6355$ \\
$\pi^+$ & weak-first pole, via $\Sig^-$        & $+D/\sqrt6$          & $+0.3282$ \\
$\pi^+$ & absorption-first pole, via $p$       & $+(D+F)/2$           & $+0.6355$ \\
$\pi^+$ & Kroll--Ruderman seagull              & $-(D+3F)/(2\sqrt6)$  & $-0.4501$ \\
$\pi^+$ & meson-in-flight ($\pi^+\to\bar K^0$) & $+(D+3F)/(2\sqrt6)$  & $+0.4501$ \\
\hline
$\bar K^0$ & line-1 absorption ($\bar K^0n\to\Lam$) & $-(D+3F)/(2\sqrt6)$ & $-0.4501$ \\
$\bar K^0$ & line-2 weak route, via $\Lam$      & $-(D+3F)/(2\sqrt6)$  & $-0.4501$ \\
$\bar K^0$ & line-2 weak route, via $\Sig^0$    & $-(D-F)/(2\sqrt2)$   & $-0.1191$ \\
$\bar K^0$ & line-2 kaon seagull               & $-(D+F)/2$           & $-0.6355$ \\
\end{tabular}
\end{ruledtabular}
\end{table}

\subsection{The contraction identities}
\label{app:cancel}

Contracting Eq.~\eqref{eq:noether3} with $q_\mu$ uses three identities, exact in the
degenerate limit. For a pole chain, $\slashed q = S^{-1}(p+q)-S^{-1}(p)$ collapses the
propagator, leaving the strong vertex evaluated on shell,
\begin{equation}
  q_\mu\,\bar u_\Lam\,\slashed k_\pi\gamma_5\,S(p_1+q)\,\gamma^\mu u_p
  = \bar u_\Lam\slashed k_\pi\gamma_5 u_p ,
\end{equation}
and analogously, with the opposite sign, for the absorption-first ordering. For the
seagull, momentum conservation at its vertex gives
\begin{equation}
  \bar u_\Lam\,\slashed q\,\gamma_5 u_p
  = 2M\,\bar u_\Lam\gamma_5 u_p - \bar u_\Lam\slashed k_\pi\gamma_5 u_p ,
\end{equation}
splitting its divergence into a $2M\gamma_5$ piece and a $\slashed k_\pi\gamma_5$ piece.
For the meson-in-flight term, $q\!\cdot\!(k_K+k_\pi) = (k_K^2-m^2)-(k_\pi^2-m^2)$ cancels
one propagator at a time, leaving one remnant proportional to the pion propagator alone,
which cancels against the pole and seagull remnants of the same exchange family, and one
proportional to the kaon propagator alone, which the kaon-exchange family cancels: its
line-2 weak route and kaon seagull carry the coefficient $(D-F)/2$ twice, once with each
sign structure. The propagator collapse requires the intermediate and external baryon
masses to coincide, and the seagull identity uses
$\bar u_Y\slashed q\gamma_5u_N = (M_N+M_Y)\bar u_Y\gamma_5u_N \to 2M\,\bar u_Y\gamma_5u_N$;
both hold in the SU(3)-symmetric limit, where the octet is degenerate, and away from it
the right-hand sides carry the physical mass splittings.

\subsection{Relations among the coefficients}
\label{app:ledger}

In the SU(3) limit the contraction identities reduce the divergence of the insertion sum
of one line assignment to four spinor--propagator structures,
\begin{equation*}
  \begin{gathered}
  D_m(k_\pi)\,\bar u_Y\slashed k_\pi\gamma_5u_{N_1}\,\bar u_{N'}\gamma_5u_{N_2},\qquad
  D_m(k_\pi)\,\bar u_Y\gamma_5u_{N_1}\,\bar u_{N'}\gamma_5u_{N_2},\\
  D_m(k_K)\,\bar u_Y\gamma_5u_{N_1}\,\bar u_{N'}\slashed k_K\gamma_5u_{N_2},\qquad
  D_m(k_K)\,\bar u_Y\slashed k_K\gamma_5u_{N_1}\,\bar u_{N'}\gamma_5u_{N_2},
  \end{gathered}
\end{equation*}
with $D_m(k)=1/(k^2-m^2)$, whose coefficients are polynomials in $D$ and $F$. They vanish,
respectively, by four relations among the coefficients of Eqs.~\eqref{eq:line2}--\eqref{eq:kexch},
each holding identically in $D$ and $F$:
\begin{align}
  \text{(L1)}\quad&\textstyle\sum_{Y_v}c^\phi_{{\rm yr},Y_v}\,f_1(N_1\!\to\!Y_v)
    -\sum_{N_v}f_1(N_v\!\to\!Y)\,c^\phi_{{\rm np},N_v}=c^\phi_{\rm sg},\nonumber\\
  \text{(M1)}\quad&c^\phi_{\rm sg}+\textstyle\sum_Kc^{\phi,K}_{\rm fl}=0,
    \label{eq:ledgerid}\\
  \text{(L2)}\quad&\textstyle\sum_{Y_v}c^K_{{\rm r},Y_v}\,f_1(N_2\!\to\!Y_v)=c^K_{\rm sg2},
    \nonumber\\
  \text{(M2)}\quad&\textstyle\sum_\phi c^\phi_2\,c^{\phi,K}_{\rm fl}=a^K_1\,c^K_{\rm sg2},
    \nonumber
\end{align}
with $f_1$ the vector coupling of each transition, Table~\ref{tab:su3}. In L1 the remnants
of the weak-first and absorption-first poles, Eqs.~\eqref{eq:hypole}
and~\eqref{eq:nucpole}, cancel the $\slashed k_\pi\gamma_5$ part of the seagull divergence;
in M1 its $2M\gamma_5$ part cancels the pion-propagator remnant of the meson in flight; in
L2 the line-2 hyperon pole of Eq.~\eqref{eq:kexch} cancels the $\slashed k_K\gamma_5$ part of the
line-2 kaon seagull; and in M2 the kaon-propagator remnants of the in-flight terms, summed
over $\phi=\pi,\eta$, cancel the $2M\gamma_5$ part of that seagull. L1 and M1 are the identity
for a single strong vertex, $W\,N_1\phi\to Y$ with the second line stripped off, and hold for
each exchanged $\phi$; L2 holds for each exchanged kaon, and
M2 for each kaon after the sum over $\phi$. All four hold within one line assignment of
one charge channel, before antisymmetrisation: each line assignment is conserved by
itself, and the $pn$ sum of Eq.~\eqref{eq:fullcurrent} and the antisymmetrised like-pair
amplitude, combinations of line-labelled amplitudes each conserved at its own kinematics,
are conserved as well. L1 and L2 follow from the Leibniz rule
for the commutator with $h$ and the completeness of the octet baryon matrices, M1 and M2
from the completeness of the meson matrices. In Table~\ref{tab:worked}, M1 pairs each
seagull with its in-flight partner, $\mp(D{+}3F)/(4\sqrt3)$ for $\pi^0$, $\mp(D{+}3F)/4$
for $\eta$ and $\mp(D{+}3F)/(2\sqrt6)$ for $\pi^+$. L1 for $\pi^0$ in assignment $A$ reads
$(D/\sqrt6)(1/\sqrt2)-\sqrt{3/2}\,(D{+}F)/(2\sqrt2)=-(D{+}3F)/(4\sqrt3)$. L2 has a single
route via $\Sig^-$ ($f_1=1$) at $(D-F)/2$ in assignment $A$, equal to the line-2 seagull,
and two routes in assignment $B$,
$-(D{+}3F)/(2\sqrt6)\cdot\sqrt{3/2}-(D{-}F)/(2\sqrt2)\cdot(1/\sqrt2)=-(D{+}F)/2$. M2 reads
$-(D{+}F)(D{+}3F)/(8\sqrt6)+(3F{-}D)(D{+}3F)/(8\sqrt6)=-(D{-}F)(D{+}3F)/(4\sqrt6)$ in
assignment $A$ and $(D{+}F)(D{+}3F)/(4\sqrt6)$ in $B$, each equal to $a_1c_{\rm sg2}$. The
equality of the absorption-first pole and line-2 emission coefficients in each table is an
isospin property of the $\pi NN$ and $\eta NN$ couplings and does not enter the closure.
The four relations hold for all ten tables of the Supplemental Material~\cite{P10supp}.
Each is necessary, and together they are sufficient: altering a single coefficient leaves
a structure uncancelled. The reference implementation returns
$|q\!\cdot\!V|/\langle|q||V|\rangle = 1.2\times10^{-15}$ per spin configuration for
assignment $A$ of this channel and $8.5\times10^{-16}$ for assignment $B$ (at most
$4.9\times10^{-15}$ for either assignment of $pn\to\Sig^0n$ and $pn\to\Sig^-p$), every
perturbation applied to it --- rescaling a single family, a single class, or a pair of
classes --- raises that residual by thirteen or more orders of magnitude, and the
production code carrying the same coefficients reproduces the full closure
(Sec.~\ref{sec:ward}). The octet route lists of all seven channels, every coefficient as
an exact combination of $D$ and $F$, are in the Supplemental Material~\cite{P10supp}.

\bibliography{refs}

\end{document}